\documentclass[aps,epsf]{revtex4}
\usepackage{graphics}
\usepackage{graphicx}
\usepackage{epsfig}

\newcommand\be{\begin{equation}}
\newcommand\ee{\end{equation}}
\def\msun{\,{\rm M_\odot}}
\def\gsim{ \lower .75ex \hbox{$\sim$} \llap{\raise .27ex \hbox{$>$}} }
\def\lsim{ \lower .75ex\hbox{$\sim$} \llap{\raise .27ex \hbox{$<$}} }

\def\apj{Astrophys.~J.}                

\def\prd{Phys.~Rev.~D }       
\def\cqg{Class.~Quant.~Grav. }

\begin{document}
\input epsf.tex

\title{Eccentric Massive Black Hole Binaries in LISA I : The Detection Capabilities of Circular Templates.}
\author{Edward K. Porter$^1$ \& Alberto Sesana$^2$\\}
\vspace{1cm}
\affiliation{$^1$APC, UMR 7164, Universit\'e Paris 7 Denis Diderot,\\ 10, rue Alice Domon et L\'{e}onie Duquet, 75205 Paris Cedex 13, France}
\affiliation{$^2$Albert Einstein Institut, 1 Am M\"uhlenberg, Golm, D-14476, Germany.}
\vspace{1cm}
\begin{abstract}
One of the major assumptions in the search for gravitational wave signatures from massive and supermassive black 
hole binaries with LISA, is that these systems will have circularized before entering the LISA bandwidth.   Current astrophysical 
simulations now suggest that systems could have a non-negligible eccentricity in the LISA band, and an important level of 
eccentricity in the Pulsar Timing regime.  In this work, we use a set of source catalogues from astrophysically motivated models of massive black hole
binary formation and assume a one year LISA mission lifetime.   Depending on the model in question, the initial eccentricities in the final year of
the inspiral can be as high as 0.6 for high mass seeds and 0.8 for low mass seeds.  We show that restricted post-Newtonian circular templates are extremely 
inefficient in recovering eccentric binaries, with median optimal signal to noise ratio recoveries of approximately 10\% for all models 
considered.  This coupled with extremely large errors in parameter recovery from individual Markov chain Monte Carlo's
demonstrate quite clearly that even to search for binaries with initial eccentricities as low as $10^{-4}$, we will require eccentric templates for LISA data analysis.

\end{abstract}

\maketitle

\section{Introduction.}
The inspiral and merger of massive and supermassive black hole binaries ((S)MBHBs) will be one of the brightest sources of gravitational waves (GW) for
the future space-borne GW detector LISA~\cite{LISA}.  There is currently a lot of effort within the community in the development of search algorithms for these
sources.  Initially, the majority of the development was conducted using non-spinning restricted post-Newtonian (PN) waveforms~\cite{cp1,cp2,cp3,bccmv,wsv,gp,fghp,bab}, i.e. the phase was constructed at a high PN order, but the amplitude was kept at the dominant Newtonian order.  In fact, these circular templates formed the basis for the initial Mock LISA data 
challenges (MLDCs)~\cite{mldc}.  More recently, people have started to investigate more complex circular waveforms such as those with higher harmonic corrections~\cite{pc,lpet,ts1,ts2,aissb} and spin~\cite{vec,lh1,lh2,mldc}.  In both cases it was comprehensively demonstrated that the added complexity helped to break parameter correlations and in some cases improved the estimation of parameters.  While these templates are definitely a step in the right direction towards using more realistic waveforms, they are still based in the circular approximation.

Until now, very little attention has been paid to the role of eccentric templates, as it had always been assumed that a binary system would circularize before 
entering the interesting frequency band~\cite{pet,pm63,mp}.  These works showed that given an initial eccentricity $e_0$ and semi-major axis $a_0$, the instantaneous eccentricity and semi-major axis were related by $e/e_0 \sim (a/a_0)^{19/12}$.  This inferred that as the semi-major axis shrunk by a factor of 2, the eccentricity shrank by a factor of almost 3.  Based on this assumption, it was shown in Ref~\cite{mp} that circular templates would be quite efficient in resolving binaries with a residual eccentricities.  However, this study made a number of assumptions that proved critical to this result.  The first was that the study assumed typical 
LIGO sources, and was focused on systems that it was assumed would (almost) circularize before reaching the lower frequency cutoff for initial LIGO at 40 Hz.  Secondly, and 
perhaps most importantly, there were no PN corrections added to the waveform.  The study used the standard Peters and Matthews equations~\cite{pet,pm63} to evolve the semi-latus rectum and eccentricity, and not the more recent PN equations for radiative dynamics.  Thirdly, the study was made for optimally orientated systems which allowed the study to neglect one of the GW polarizations.

More recently, two separate studies in particular have shown the previous results to be erroneous~\cite{TG, BZ}.  These particular studies included PN corrections in both the conservative dynamics and radiation reaction.  The overwhelming conclusion of both of these studies was that circular binaries were incapable of matching eccentric systems once the initial eccentricity was greater than 0.1.  We should point out, that these studies were again based on a LIGO study where the waveform durations are very short (on the timescale of seconds).  However it was highlighted in Ref~\cite{BZ} that longer duration signals, such as neutron star - neutron star binaries had lower matches with eccentric binaries.  This was due to the fact that there is more of a possibility of de-phasing between waveforms if the signal is of long enough duration.  It is for this very reason that it is important to conduct this investigation for LISA, where signals have durations of months to years. 

Furthermore, although the circular orbit ansatz has been widely used by the GW community, the reliability of such an assumption
is now further questionable from an astrophysical point of view. Since the evolution of MBHBs was firstly sketched by Begelman Blandford \& Rees \cite{br80}, hundreds of studies have been 
dedicated to the subject. After the two MBHs, driven by dynamical friction \citep{bt87,cmg99}, 
reach the center of the merged system 
and pair together, the new-formed binary needs to transfer energy and angular momentum to the surrounding ambient in order to coalesce. Typical 
pairing lengthscales of {\it LISA} binaries (binary masses in the range $10^4-10^7\msun$) 
are of the order of 0.1-1 pc; however GW emission is efficient in driving
binaries to the final coalescence only at mpc scales. This pc-to-mpc gap in the MBHB evolution
goes under the name of 'last parsec problem' \citep{mm01}. 
In stellar environments, the MBHB evolution proceeds via super-elastic 
scattering of surrounding stars intersecting the binary orbit \citep[slingshot
mechanism, ][]{mv92}, and the fate of the system depends on the supply of stars
available for such interaction \citep{yu02,mp04,ms06,pa08}. On the other hand, if the system is gas
rich, torques exerted by a massive circumbinary disk have been proven 
efficient in shrinking the binary down to $\sim0.1$ pc \citep{escala05,dotti07},
which is the current resolution limit of dedicated smoothed particle
hydrodynamical simulations. However, whether viscous angular momentum extraction is efficient 
all the way down to coalescence is questionable \citep{lod09}.
Although there are still open questions about the effectiveness of
these dynamical processes \citep{mm03}, both stellar and gas based shrinking mechanisms have 
proven to be efficient in increasing the binary eccentricity
\citep{qui96,an05,shm06,shm08,bau06,mat07,ber09,cua09,as09,as10}.
Whether a MBHB is going to be circular or eccentric for GW detection purposes, depends on
the amount of 'residual eccentricity' left by GW-driven circularization by the time the system 
enters the LISA window. Using an hybrid model to couple scattering of bound and unbound
stars and GW emission, Sesana \cite{sesana10} carried a systematic study of the residual eccentricity
of MBHBs evolving in stellar environments. Assuming standard MBH formation and evolution 
scenarios \citep{vhm03,bvr06} he demonstrated that  LISA MBHBs are likely to show a significant
amount of eccentricity, calling for the development of trustworthy eccentric templates for
MBHB detection and parameter estimation. 

In this paper we test the capability of restricted PN circular templates in detecting GWs from eccentric MBHBs for LISA.  As well as
detection capabilities, we are also interested in any parameter mismatch as a result of the inclusion of eccentricity.  Furthermore, there has also been 
incredible progress made in the field of numerical relativity (NR)~\cite{nr1, nr2, nr3} in the last number of years, providing the GW community with catalogues of merger waveforms for both non-spinning and spinning black hole binaries.  However, a priority in the field has been to find ways of reducing the residual eccentricity in the NR waveforms~\cite{re, re2}.  It will be interesting to see what our results mean for both of these endeavors.  

 For this study, we use catalogs of eccentric MBHBs evolved according to the scheme presented in \cite{sesana10}.  This provided us with six individual catalogues as our
starting point.  For each model, the binary reaches a separation where dynamical friction is no longer efficient in shrinking the binary orbit.  As the dynamics leading to this point are highly environment dependent, the eccentricity at this point has to be added by hand at the beginning of simulations.  Thus,  the catalogues were generated
for both high and low mass seed black hole binaries with initial eccentricities of $e_0=0, 0.3$ and 0.6.  

The paper is organized as follows. In Section~\ref{sec:lisa} we present the response of the LISA detector in the low frequency approximation.  Section~\ref{sec:waveforms}
contains a description of the waveforms for both circular and eccentric binary systems.  We then describe the astrophysical model used to produce the source
catalogues in Section~\ref{sec:model}, before describing the setup of the Monte Carlo simulation in Section~\ref{sec:mc}.  In Section~\ref{sec:results} we present
the results of the Monte Carlo runs.  

Throughout the paper we use the units $G=c=1$.

\section{The LISA Detector Response In the Low Frequency Approximation.}\label{sec:lisa}
The response $h(t)$ of the A and E LISA channels to an incoming GW with polarizations $h_+$ and $h_\times$ in the low frequency approximation~\cite{cutler} is given by the combination 
\begin{equation}
h_{A,E}(t) = h_{+}(\xi(t))F^{+}_{A,E}(t)+h_{\times}(\xi(t))F^{\times}_{A,E}(t), 
\end{equation}
where the phase shifted time parameter $\xi(t)$ is defined by
\begin{equation}
\xi(t) = t - R_{\oplus}\sin\theta\cos\left(\alpha(t) - \phi\right).
\end{equation}
Here $R_{\oplus}$ denotes 1 AU, $(\theta,\phi)$ are the sky location of the system and $\alpha(t) = 2\pi f_m t + \kappa$, where $f_m$ is the LISA modulation frequency and $\kappa$ is the longitudinal offset of the detector array.  The polarizations of the GW $h_{+,\times}(t)$ will be defined at later stage.
The beam pattern functions are defined by
\begin{equation}
F^{+}_{A,E}(t) = \frac{1}{2}\left[\cos(2\psi)D^{+}_{A,E}(t;\theta, \phi, \lambda) - \sin(2\psi)D^{\times}_{A,E}(t;\theta, \phi, \lambda)\right],
\end{equation}
\begin{equation}
F^{\times}_{A,E}(t) = \frac{1}{2}\left[\sin(2\psi)D^{+}_{A,E}(t;\theta, \phi, \lambda) + \cos(2\psi)D^{\times}_{A,E}(t;\theta, \phi, \lambda)\right].
\end{equation}
The quantity $\psi$ is the polarization angle of the wave.  Formally, if ${\bf \hat{L}}$ is the direction of the binary's orbital angular momentum and ${\bf\hat{n}}$ is the direction from the observer to the source (such that the GWs propagate in the ${-\bf\hat{n}}$ direction), then $\psi$ fixes the orientation of the component of ${\bf\hat{L}}$ perpendicular to ${\bf\hat{n}}$.  The time dependent quantities $D_{+,\times}(t)$ are given in the LFA by~\cite{cornishrubbo} 
\begin{eqnarray}
D^{+}_{A,E}(t)& =& \frac{\sqrt{3}}{64}\left[\frac{}{}-36\sin^{2}(\theta)\sin(2\alpha(t)-2\lambda)+(3+\cos(2\theta)) \right.\\ \nonumber&& \left(\frac{}{}\cos(2\phi)\left\{\frac{}{}9\sin(2\lambda)-\sin(4\alpha(t)-2\lambda)\right\} \frac{}{}+\sin(2\phi)\left\{\frac{}{}\cos\left(4\alpha(t)-2\lambda\right)-9\cos(2\lambda) \right\}\frac{}{}\right)\\ \nonumber  &&\left.-4\sqrt{3}\sin(2\theta)\left(\frac{}{}\sin(3\alpha(t)-2\lambda-\phi)-3\sin(\alpha(t)-2\lambda+\phi)\right)\right]\\
D^{\times}_{A,E}(t)& = &\frac{1}{16}\left[\frac{}{}\sqrt{3}\cos(\theta)\left(\frac{}{}9\cos(2\lambda-2\phi)-\cos(4\alpha(t)-2\lambda-2\phi) \right) \right. \\ \nonumber &&\left. -6\sin(\theta)\left(\frac{}{} \cos(3\alpha(t)-2\lambda-\phi)+3\cos(\alpha(t)-2\lambda+\phi) \right) \right],
\end{eqnarray}
where $\lambda = 0$ and $\pi/4$ give the orientation of the two detectors, thus defining the A and E channels.  To ensure that we are working in the LFA domain, we limit the waveforms to a GW frequency of $f_{gw}(6m)$ or 5 mHz, whichever is lower.

\section{The Gravitational Waveform From Circular and Eccentric Binaries.}\label{sec:waveforms}
For this study, we use circular and eccentric non-spinning waveforms.   For the circular templates, we use a waveform with 2 PN corrections to both the orbital phase and angular frequency evolution.  For the eccentric waveforms we assume 2 PN corrections to both the conservative and adiabatic dynamics of the system.  To make the comparison with the eccentric waveforms, we are using the standard restricted PN circular templates that have been used extensively in other studies.  The reason for this is, we would like to study the fidelity of the non-spinning circular templates that have been adopted by the community in capturing the eccentric systems, rather than using the zero eccentricity limit of the eccentric waveforms.  We will therefore describe each waveform in turn.

\subsection{The Circular Binary Waveform}
The GW polarizations for a non-spinning circular binary are given by
\begin{eqnarray}
h_{+}& =& \frac{2m\eta}{D_{L}}\left(1+\cos^{2}(\iota)\right)x\cos\left(2(\varphi_c - \Phi)\right),\\ \nonumber \\
h_{\times} &= &-\frac{4m\eta}{D_{L}}\cos(\iota)\,x\sin\left(2(\varphi_c -\Phi)\right).
\end{eqnarray}
Here $m=m_{1}+m_{2}$ is the total mass of the binary, $\eta = m_{1}m_{2}/m^{2}$ is the reduced mass ratio and $D_L$ is the luminosity distance, which is related to the redshift $z$ of the source by
\begin{equation}
D_{L} = \frac{c(1+z)}{H_{0}}\int_{0}^{z}\,dz'\left[\Omega_{R}\left(1+z'\right)^{4}+\Omega_{M}\left(1+z'\right)^{3}+\Omega_{\Lambda}\right]^{-1/2}, 
\end{equation}
where we use the WMAP values of $(\Omega_{R}, \Omega_{M}, \Omega_{\Lambda}) = (4.9\times10^{-5}, 0.27, 0.73)$ and a Hubble's constant of $H_{0}$=71 km/s/Mpc~\cite{wmap}.  The inclination of the orbit of the binary system is formally defined as $\cos\iota=\bf\hat{L}\cdot\hat{n}$.  The quantity $\varphi_c$ is the orbital phase constant at coalescence and the invariant PN velocity parameter is defined by $x(t) = \left(m\omega(t)\right)^{2/3}$, where 
\begin{eqnarray}\label{eqn:freq}
\omega(t)=\frac{1}{8m}\left[\Theta^{-3/8}+\left(\frac{743}{2688}+\frac{11}{32}\eta\right)\Theta^{-5/8}-\frac{3\pi}{10}\Theta^{-3/4}+\left(\frac{1855099}{14450688}+\frac{56975}{258048}\eta+\frac{371}{2048}\eta^{2}\right)\Theta^{-7/8}\right],
\end{eqnarray}
is the 2 PN order orbital angular frequency for a circular orbit formally defined as $\omega=d\Phi/dt$, and $\Phi$ is the orbital phase which is given by
\begin{eqnarray}\label{eqn:phase}
\Phi(t) = \frac{1}{\eta}\left[\Theta^{5/8}+\left(\frac{3715}{8064}+\frac{55}{96}\eta\right)\Theta^{3/8}-\frac{3\pi}{4}\Theta^{1/4}+\left(\frac{9275495}{14450688}+\frac{284875}{258048}\eta+\frac{1855}{2048}\eta^{2}\right)\Theta^{1/8}\right].
\end{eqnarray}
We should note that the gravitational wave phase is defined by $\Phi_{GW} = 2\Phi$.  The time dependent quantity $\Theta(t;t_{c})$ is related to the time to coalescence of the wave, $t_{c}$, by
\begin{equation}
\Theta(t;t_{c}) = \frac{\eta}{5m}\left(t_{c}-t\right).
\end{equation}
As the PN waveforms are known to break down in certain cases before we reach $r=6m$, we use a taper function to smoothly truncate the waveform.

\subsection{The Eccentric Binary Waveform.}
The restricted 2 PN waveform polarisations for non-spinning eccentric binaries are given by
\begin{eqnarray}
h_+(t) &= &\frac{m\eta}{ D_L}  \left[\left(1+\cos ^2(\iota) \right) \bigg[\cos \left(2\left(\varphi_c - \Phi\right) \right)  \left(-\dot{r}^2+r^2 \dot{\Phi   }^2+\frac{m}{r}\right)     +2 r\, \dot{r}\, \dot{\Phi } \sin \left(2( \varphi_c - \Phi)\right) \bigg ] \right. \\   
&+&\left.\left(-\dot{r}^2-r^2 \dot{\Phi }^2+\frac{m}{r}\right) \sin ^2 (\iota) \right]  \nonumber \\ 
h_\times(t) &=& -\frac{2 m \eta}{ D_L}  \cos( \iota)  \left[\left(-\dot{r}^2+r^2  \dot{\Phi }^2+\frac{m}{r}\right) \sin \left(2( \varphi_c - \Phi)\right)   -2 r\,\dot{r}\, \dot{\Phi } \cos \left(2 (\varphi_c - \Phi)\right) \right]\, .\\ \nonumber
\end{eqnarray}
The components $(r(t), \Phi(t))$ denote the orbital seperation and phase of the system (also referred to as the true anomaly), and $\dot{r}(t)=dr/dt$ and $\dot{\Phi}(t)=d\Phi/dt$.  We noted that in the circular binary case, the angular frequency $\omega$ is defined by $\omega=d\Phi/dt$.  This is not true for eccentric binaries and furthermore, $d\Phi/dt$ is no longer a monotonic function of time.   We will however continue to define the PN velocity parameter $x$ as $x=(m\omega)^{2/3}$, but now where $\omega\equiv(2\pi +\Delta\Phi)/P = n+\Delta\Phi/P$.  Here $n$ is the mean motion of the binary system and $P=2\pi/n$ is the radial orbital period.  We should point out here that $P$ is defined as the time to go from pericenter to pericenter.  Due to precession effects, this is different from the time taken to go from $\Phi$ to $\Phi+2\pi$.  The parameter $\Delta\Phi$ represents the advance of the pericenter per period.   It was shown in Ref~\cite{TG2} that while the dominant spectral component for a circular binary appears at $f_{gw} = n/\pi$, this changes to $(1+k)n/\pi$ for an eccentric binary having a PN accurate orbital motion, where $k = \Delta\Phi/2\pi$.

Previous works evolve the eccentric binary in terms of the parameters $(n,e)$~\cite{TG, TG2, GI, DGI, KG, ABIQ}, where again $n$ is the mean motion and $e$ is the eccentricity of the system.  However, Hinder et al~\cite{hinder} demonstrated that using the parameters $(x,e)$ provides a better match with waveforms from numerical relativity.  For this reason, we will also work with the parameter pair $(x, e)$.  To go from $(n,e)$ to $(x,e)$ we use the following 2 PN relation
\begin{equation}
mn = x^{3/2} + \frac{3}{e^2 - 1} x^{5/2} + \frac{(26\eta-51)e^2+28\eta-18}{4(e^2 - 1)^2} x^{7/2} .
\end{equation}

It was shown that eccentricity effects the waveform of a SMBHB in three ways~\cite{BZ}.  First of all, the eccentricity induces amplitude modulation, secondly, it increases the amplitude of the waveform, and thirdly, it decreases the duration of the signal.

\subsubsection{The PN Conservative Orbital Dynamics.}
In order to write the relevant quantities in terms of $(x,e)$, we substitute the right hand side of the above equation into the necessary expressions in~\cite{KG} which are expressed  in terms of $(n,e)$ and truncate the subsequent equations at the 2 PN order.  The parameters describing the conservative orbital dynamics can now be described in terms of the quantities $(x(t), e(t), u(t))$
\begin{eqnarray}
r(t)/m &=& \left(1-e \cos (u)\right)x^{-1} + \left[\frac{2 (e \cos (u)-1)}{e^2-1}+\frac{1}{6} (2 (\eta -9)+e (7 \eta -6) \cos (u))\right] \\ \nonumber
&+& \left[\frac{e^2(51-26\eta)-28\eta+48}{(e^2 -1)^2}(1-e\cos(u)) + \frac{1}{72(1-e^2)}\left\{504\eta-288-36(5-2\eta)(2+e\cos(u))\sqrt{1-e^2} \right. \right. \\ \nonumber
&+&\left.\left.\left[72+30\eta+8\eta^2 - (72-231\eta+35\eta^2) e\cos(u)\right](1-e^2)\frac{}{}\right\}\right] x , \\ \nonumber \\ \nonumber \\ 
\dot{r}(t) & = & \frac{e\sin(u)}{(1-e\cos(u))}\left[x^{1/2} + \frac{1}{(e^2 - 1)}x^{3/2} + \frac{e^2(26\eta-51)+28\eta-30}{(e^2 -1)^2}x^{5/2}     \right], \\ \nonumber \\ \nonumber \\
\Phi(t) & = & \left(1+\frac{3}{1-e^2}x + \frac{54-28\eta + e^2(61-26\eta)}{4(e^2 - 1)^2}x^2\right) l + (v-u) + e\sin(u) + \frac{3(v-u+e\sin(u))}{1-e^2}x\\ \nonumber
&+&\left[\frac{6(v-u+e\sin(u))}{(e^2 -1)(1-e^2)} + \frac{1}{32(1-e^2)^2(1-e\cos(u))^3}\left[ 8\left[78-28\eta+(51-26\eta)e^2 - 6(5-2\eta(1-e^2))^{3/2}\right]  \right.\right.\\ \nonumber\\ \nonumber
&\times&(v-u)(1-e\cos(u))^3+\left[624-284\eta+4\eta^2+(408-88\eta-8\eta^2)e^2 - (60-4\eta)\eta e^4  \right.\\ \nonumber
&+& \left\{792\eta-1872-8\eta^2-(1224-384\eta-16\eta^2)e^2 + (120-8\eta)\eta e^4\right\}e\cos(u) + \left\{ 1872-732\eta+4\eta^2  \right.\\ \nonumber
&+&\left. \left.\left(1224-504\eta-8\eta^2\right)e^2 - \left(60-4\eta\right)\eta e^4\right\} (e\cos(u))^2 + \left\{224\eta-624-(408-208\eta)e^2   \right\}(e\cos(u))^3\right]\\ \nonumber
&\times&e\sin(u) + \left[(27\eta^2 - 153\eta-8)e^2 + (4\eta-12\eta^2)e^4+\left\{8+152\eta-24\eta^2+(8+146\eta-6\eta^2)e^2   \right\}e\cos(u)  \right.\\ \nonumber
&+&\left.\left.\left\{12\eta^2-148\eta-8-(\eta-3\eta^2)e^2 \right\}(e\cos(u))^2\right] e\sin(u)\sqrt{1-e^2}\frac{}{}\right]x^2, \\ \nonumber \\ \nonumber \\
m\dot{\Phi}(t) &=& \frac{\sqrt{1-e^2}}{(1-e\cos(u))^2}x^{3/2}+\frac{e(e-\cos(u))(4-\eta)}{\sqrt{1-e^2}(e\cos(u)-1)^3}x^{5/2}+\left[\frac{1}{12(1-e^2)^{3/2}(e\cos(u)-1)^5}\left\{\frac{}{}\left[(42+22\eta+8\eta^2) \right. \right. \right.\\ \nonumber
&+& \left.e^2(8\eta-147-14\eta^2) + \sqrt{1-e^2}(36\eta-90)\right](e\cos(u))^3+\left[ e^4(48+17\eta^2-17\eta)+e^2(153-38\eta-4\eta^2) \right.\\ \nonumber
&+&\left.(5\eta^2+114-35\eta)+\sqrt{1-e^2}(e^2(180-72\eta)-36\eta+90)\right](e\cos(u))^2+\left[e^4(12+97\eta-\eta^2) \right.\\ \nonumber
&-&\left.e^2(81+74\eta+16\eta^2)+(67\eta-246-\eta^2)+\sqrt{1-e^2}(e^2(144\eta-360)+90-36\eta)\right]e\cos(u)\\ \nonumber
&+&\sqrt{1-e^2}(e^2(180-72\eta)+36\eta-90)-e^6(12\eta^2-18\eta)+e^4(26\eta+20\eta^2-60)\\ \nonumber
&+&\left.\left. e^2(75+50\eta-2\eta^2)-36\eta +90\frac{}{}\right\}\right]x^{7/2}.\nonumber
\end{eqnarray}
In the above equations, $l(t)$ is the mean anomaly and is found by integrating the following identity
\begin{equation}\label{eqn:dldt}
m\frac{dl}{dt} = mn = x^{3/2} + \frac{3}{e^2 - 1} x^{5/2} + \frac{(26\eta-51)e^2+28\eta-18}{4(e^2 - 1)^2} x^{7/2} .
\end{equation}
The quantity $u(t)$ is the eccentric anomaly and is determined by solving the transcendental 2 PN Kepler equation
\begin{equation}\label{eqn:uoft}
l = u-e\sin(u) + \frac{1}{8\sqrt{1-e^2}(1-e\cos(u))}\left[-12(2\eta-5)(u-v)(e\cos(u)-1)-e\sqrt{1-e^2}(\eta-15)\eta\sin(u)\right]x^2,
\end{equation}
where the term $(u-v)$ is defined as
\begin{equation}
v-u = 2\tan^{-1}\left(\frac{\sin(u)\beta_{\phi}}{1-\cos(u)\beta_{\phi}}\right),
\end{equation}
$\beta_{\phi}$ is given by
\begin{equation}
\beta_{\phi} = \frac{1-\sqrt{1-e_{\phi}^2}}{e_{\phi}},
\end{equation}
and
\begin{eqnarray}
e_{\phi} = e\left(1-(\eta-4)x+\frac{x^2}{96(e^2 -1)}\left[ (41\eta^2 - 659\eta+1152)e^2 + 4\eta^2 + 68\eta+\sqrt{1-e^2}(288\eta-720)-1248  \right]\right).
\end{eqnarray}
We should point here that one of the differences in the evolution of the orbital phase between circular and eccentric waveforms, is that part of the secular evolution
of $\Phi(t)$ for an eccentric binary is due to advance of periastron.  This effect is taken into account by the leading term in the expression for $\Phi(t)$ given above. 

\subsubsection{The PN Radiation Reaction}
To describe the effects of radiation reaction, the adiabatic evolution of $x(t)$ and $e(t)$ are given by the set of coupled 1st order differential equations
\begin{eqnarray}
m \dot x &=& \frac{2 \big(37 e^4+292 e^2+96\big) \eta }{15 \big(1-e^2\big)^{7/2}} x^5 + \frac{\eta}{420(1-e^2)^{7/2}}\left[-(8288 \eta -11717) e^6-14 (10122 \eta -12217) e^4\right.\\ \nonumber
&-&\left.120 (1330 \eta -731) e^2-16 (924 \eta +743) \right] x^{6} + \frac{256}{5}\eta\kappa_E(e)x^{13/2} + \frac{\eta}{45360(1-e^2)^{11/2}}\left[\big(1964256 \eta ^2\right.\\ \nonumber
&-&\left.3259980 \eta +3523113\big) e^8+\big(64828848 \eta ^2-123108426 \eta +83424402\big) e^6+\big(16650606060 \eta ^2-207204264 \eta \right.\\ \nonumber
&+&\left.783768\big) e^4+\big(61282032 \eta ^2+15464736 \eta -92846560\big) e^2+1903104 \eta ^2+\sqrt{1-e^2} \big((2646000-1058400 \eta ) e^6\right. \\ \nonumber
&+&\left.(64532160-25812864 \eta ) e^2-580608 \eta +1451520\big)+4514976 \eta -360224\right] x^{7} , 
\end{eqnarray}
\begin{eqnarray}
m \dot e &=& -\frac{e \big(121 e^2+304\big) \eta }{15 \big(1-e^2\big)^{5/2}}  x^4 + \frac{e \eta }{2520 \big(1-e^2\big)^{7/2}}\left[ \frac{}{}(93184 \eta -125361) e^4+12 (54271 \eta -59834) e^2\right.\\ \nonumber
&+&\left.8 (28588 \eta +8451)\frac{}{}\right] x^{5} +  \frac{128\eta\pi}{5e} \left[\big(e^2-1\big) \kappa _E(e)+\sqrt{1-e^2} \kappa _J(e) \right] x^{11/2} - \frac{e\eta}{30240(1-e^2)^{9/2}}\left[\frac{}{}13509360\eta   \right.\\ \nonumber
&-& 15198032 + \big(2758560 \eta ^2-4344852 \eta +3786543\big) e^6+\big(42810096 \eta ^2-78112266 \eta +46579718\big) e^4\\ \nonumber
&+&\big(48711348 \eta ^2-35583228 \eta -36993396\big) e^2+4548096 \eta ^2+\sqrt{1-e^2} \big((2847600-1139040 \eta ) e^4\\ \nonumber
&+&\left.(35093520-14037408 \eta ) e^2-5386752 \eta +13466880\big)\frac{}{}\right]x^6.
\end{eqnarray}
These quantities provide the dominant secular evolution for the conservative orbital parameters $(r, \dot{r}, \Phi, \dot{\Phi}, l)$.  We can see that the above equations are functions of $(x,e)$ only.  In order to generate the polarizations of the waveform, we first evolve this set of coupled ODEs.  Once we have $(x(t), e(t))$, we can integrate Eqn~(\ref{eqn:dldt}) for $l(t)$, and then solve the transcendental Kepler Equation, i.e. Eqn~(\ref{eqn:uoft}), for $u(t)$.  We then have all necessary quantities to then evolve the conservative orbital parameters.

The two functions $\kappa_E(e)$ and $\kappa_J(e)$ which appear in the expressions for the tail terms at the 1.5 PN order in $(\dot{x}, \dot{e})$ are defined in terms of the following infinite series
\begin{eqnarray}
\kappa_E &=& \sum_{p=1}^{\infty} \frac{1}{4}p^3\left[\left\{ \left(-e^2 - \frac{3}{e^2} + \frac{1}{e^4} + 3\right)p^2 + \frac{1}{3} - \frac{1}{e^2} + \frac{1}{e^4} \right\}J_{p}^{2}(pe) + \left(-3e - \frac{4}{e^3} + \frac{7}{e}\right) p J'_{p}(pe) J_{p}(pe)  \right. \\ \nonumber
&+& \left.\left\{\left(e^2+\frac{1}{e^2} - 2\right)p^2 + \frac{1}{e^2} - 1\right\}(J'_{p}(pe))^2\right],
\end{eqnarray}
and
\begin{eqnarray}
\kappa_J &=& \sum_{p=1}^{\infty}\frac{1}{2}p^2\sqrt{1-e^2}\left[ \left(-\frac{2}{e^4} -1 + \frac{3}{e^2} \right)p  J_{p}^{2}(pe) + \left\{2\left(e + \frac{1}{e^3} - \frac{2}{e} \right)p^2 - \frac{1}{e} + \frac{2}{e^3}\right\}J'_{p}(pe) J_{p}(pe) \right. \\ \nonumber
&+& \left. 2\left(1-\frac{1}{e^2}\right)p(J'_{p}(pe))^2 \right],
\end{eqnarray}
where $J_p(pe)$ are Bessel functions of the first kind, and $J'_{p}(pe)$ is given by
\begin{equation}
J'_{p}(pe) = J_{p-1}(pe)-\frac{J_p(pe)}{e}.
\end{equation}
Rather than evaluate the summations themselves, we expanded both expressions in terms of truncated shifted Chebyshev series  
\begin{equation}
\kappa_{E,J} \approx \sum_{k=0}^{n} \alpha_k\,T^{*}_{k}(y)
\end{equation}
where $T^{*}_k(y)$ denotes the shifted Chebyshev polynomial over the domain $10^{-5}\leq e \leq 0.6$, and where 
\begin{equation}
y = \frac{ 200000}{59999}e-\frac{60001}{59999}.
\end{equation}
It was shown in~\cite{porcheb} that the truncation error in a shifted Chebyshev series is equal to the coefficient of the truncated term.  In this case, demanding a maximum error of $10^{-10}$ over the required interval, the infinite series for $\kappa_E(e)$ and $\kappa_J(e)$ reduce to the following finite shifted Chebyshev series
\begin{eqnarray}
\kappa_E& =& \frac{6089404953}{627737060}T^{*}_0(y)+\frac{7361480108}{511018171}T^{*}_1(y)+\frac{4082990728}{477749665}T^{*}_2(y)+\frac{3631828436}{883322963}T_3(y)+ \frac{12384069875}{6968532396}T^{*}_4(y) \\ \nonumber \\ \nonumber
&+& \frac{3874479909}{5563217669}T^{*}_5(y) + \frac{200275816}{781767297}T^{*}_6(y) + \frac{1411569112}{15834804511}T^{*}_7(y) + \frac{691406503}{23234918247}T^{*}_8(y) + \frac{652605463}{68094099916}T^{*}_9(y) \\ \nonumber \\ \nonumber
&+&  \frac{165805721}{55300342056}T^{*}_{10}(y) + \frac{307444903}{336107341116}T^{*}_{11}(y) + \frac{15533911}{56864529638}T^{*}_{12}(y) + \frac{31472697}{393051374845}T^{*}_{13}(y) \\ \nonumber \\ \nonumber
&+&  \frac{10615571}{459713375092}T^{*}_{14}(y) + \frac{7366315}{1122247921676}T^{*}_{15}(y) + \frac{3467836}{1882611578695}T^{*}_{16}(y) + \frac{1861768}{3643208150105}T^{*}_{17}(y) \\ \nonumber  \\ \nonumber
&+&  \frac{1673748}{11928896825275}T^{*}_{18}(y) + \frac{772027}{20228638986003}T^{*}_{19}(y) + \frac{1309855}{127258169138523}T^{*}_{20}(y) +  \frac{97229}{35299737783375}T_{21}(y)\\ \nonumber  \\ \nonumber
&+&\frac{86276}{117891810143169}T^{*}_{22}(y),
\end{eqnarray}
and
\begin{eqnarray}
\kappa_J &=& \frac{4295041741}{1253808123}T^{*}_{0}(y)+\frac{9397451575}{2480459309}T^{*}_{1}(y)+\frac{5904096201}{3128677709}T^{*}_{2}(y)+\frac{1680140797}{2332139314}T^{*}_{3}(y)+\frac{249444067}{954292578}T^{*}_{4}(y) \\ \nonumber \\ \nonumber
&+&\frac{922739783}{10740607462}T^{*}_{5}(y)+\frac{970838387}{35705967112}T^{*}_{6}(y)+\frac{405641150}{49419308909}T^{*}_{7}(y)+\frac{121415427}{50361971216}T^{*}_{8}(y)+\frac{148577254}{215643317571}T^{*}_{9}(y) \\ \nonumber \\ \nonumber
&+&\frac{52504253}{271988209200}T^{*}_{10}(y)+\frac{9209709}{173392053574}T^{*}_{11}(y)+\frac{10429485}{724157789068}T^{*}_{12}(y)+\frac{3461427}{897971624647}T^{*}_{13}(y) \\ \nonumber \\ \nonumber
&+&\frac{698473}{684567257610}T^{*}_{14}(y)+\frac{1150943}{4302914322681}T^{*}_{15}(y)+\frac{1609829}{23072557955755}T^{*}_{16}(y)+\frac{756774}{39666328924501}T^{*}_{17}(y) \\ \nonumber \\ \nonumber
&+&\frac{152127}{16586813958911}T^{*}_{18}(y).
\end{eqnarray}
We have also verified that outside of these intervals, the Cheybshev series have maximum errors of $10^{-8}$.  To evaluate the shifted Chebyshev polynomials, we can use the fact that
\begin{equation}
T_{0}^{*}(y)=1 ,\,\,\,\,\,\,\,\,\,\,T_{1}^{*}(y) = y,
\end{equation}
which now allow us to use the recurrence relation
\begin{equation}\label{eqn:reqeqn2}
T_{k}^{*}(y)=2y\,T_{k-1}^{*}(y)-T_{k-2}^{*}(y)\,\,\,\,\,\,\,\,\,\,\,\,\,k\geq 2,
\end{equation}
to calculate the higher order shifted polynomials.

We should point out that in its formal presentation,  the above waveform description does not have support at $e\equiv 0$.  This can be seen from the fact that at $e = 0$, $\beta_{\phi}  \equiv \infty$ and $(u-v)$ is undefined.  On the other hand, $\kappa_E$ and $\kappa_J$ asymptotically approach unity as $e\rightarrow0$, but again are undefined at $e\equiv0$ due to infinities appearing in the expressions.  We should also mention that at values of eccentricity of $e\leq10^{-7}$ the formal expressions for $\kappa_E$ and $\kappa_J$ become numerically unstable and begin to oscillate around unity.  While our goal was always to compare the circular templates that have been used in the literature with eccentric templates, it is for these reasons that we need to use specifically circular templates and not just set $e=0$ in the eccentric templates.

\section{Astrophysical Modeling of Low and High Mass Seed Black Hole Binaries.}\label{sec:model}
To produce trustworthy catalogs of coalescing eccentric MBHBs, we proceed in two
steps. Firstly, we extract catalogs of MBHB masses and redshifts from
standard models of hierarchical MBH formation and evolution; then, we track the
eccentricity of each system applying an hybrid model for the binary
evolution in stellar dominated environments.

\subsection{Cosmological population of massive black holes}
MBHs are a ubiquitous components of nearby galaxy nuclei  
\citep[see, e.g., ][]{mago98}, and their masses tightly correlate with
the properties of the host \citep[][and reference therein]{hr04}. 
In popular $\Lambda$CDM cosmologies, structure formation proceeds
in a hierarchical fashion \citep{wr78}, in which massive galaxies  
are the result of several merging events involving smaller building blocks.
In this framework, the MBHs we see in galaxies today are expected to be the natural 
end-product of a complex evolutionary path, in which black holes
seeded in proto-galaxies at high redshift grow along the cosmic
history through a sequence of mergers and accretion episodes \citep{kh00,vhm03}.
Hierarchical models for MBH evolution, associating quasar activity to gas-fueled
accretion following mergers between galaxies, have been successful in reproducing
several properties of the observed Universe, such as the present day mass density
of nuclear MBHs and the optical and X-ray luminosity functions of quasars 
\citep{vhm03,m07}. 

\begin{figure}
\centering
\includegraphics[width=0.55\linewidth]{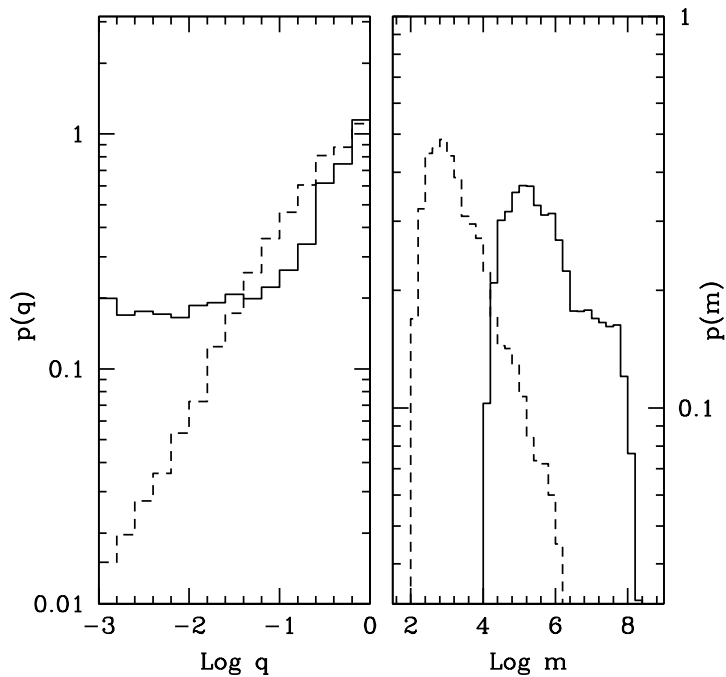}
\caption{Mass (right panel) and mass ratio (left panel) distribution 
of coalescing MBHBs according to our MBH evolution models. The solid
and the dashed histograms refer to the heavy and the light seed model
respectively. The histograms are normalized so that the integrals 
in $d{\rm log}q$ and $d{\rm log}m$ are unity.}
\label{fdist}
\end{figure}

In this general picture, the mechanism responsible for the formation of the first 
seed BHs is not well understood, and two distinctive families of models 
have became popular in the last decade.
In the first family, seeds are light 
\citep[$M\gsim100\msun$, 'light seed' scenario][]{vhm03}, 
being the remnant of the first POPIII star explosions \citep{mr01}; 
in the second one, already quite heavy ($M\gsim10^4\msun$) seed BHs form by
direct collapse of massive proto-galactic discs 
\citep['heavy seed' scenario][]{bvr06,kbd04}.
The two models adopted here, representative of the 
two scenarios, are the same employed by the {\it LISA} parameter estimation
task force \citep{lpet} (for the reader interested in more details,
our light and heavy seed models correspond to models LE and SE in \citep{aru09},
respectively). Each model is constructed tracing backwards
the merger hierarchy of 220 dark matter halos in the mass
range $10^{11}-10^{15}\msun$  up to z = 20, using an extended
Press \& Schechter (EPS) algorithm (see \citep{vhm03} for details). 
The halos are then seeded with seed black holes and their 
evolution is tracked to the present time. Following a major merger
(defined as a merger between two halos with mass ratio $M_2/M_1>0.1$,
being $M_2$ the mass of the lighter halo) 
MBHs accrete efficiently, at the Eddington rate, an amount of mass 
that scales with the fifth power of the host halo circular velocity, normalized
to reproduce the observed local correlation between MBH mass
and the bulge stellar velocity dispersion ($M-\sigma$ relation, see
\citep{tr02} and references therein). 
For each of the 220 halos, 
all the coalescence events happening during the cosmic history are collected.
The outputs are then weighted using the EPS halo mass function
and integrated over the observable volume shell at every redshift to
obtain numerically the coalescence rate of MBHBs as a function of
black hole masses and redshift (see, e.g., Fig. 1 in \citep{svh07}). 
In other words, the outcome of
this procedure is the numerical distribution $d^4N/dzdm_1dm_2dt$.
We then perform 1000 Monte Carlo sampling of the $d^4N/dzdm_1dm_2dt$
generated by each model, producing 1000 catalogues of coalescing binaries 
over a period of one year. Distribution of MBH masses and mass ratios predicted
by the two models are plotted in figure \ref{fdist}. The mass distribution
is peaked around $10^3\msun$ for the light seed scenario, and around
$10^5\msun$ for the heavy seed scenario. Both mass ratio distributions are 
peaked at $q=m_2/m_1=1$, however, the low mass seed model has a more gentle
behavior, and coalescences are spread in the mass ratio range $0.1-1$. 

\subsection{Dynamics of eccentric massive black hole binaries in stellar environments}
The next step is to attach to each binary in the catalog an orbital
eccentricity at some selected frequency, relevant for GW observations.
Here is the description of our methodology. 
Following a major merger, the two MBHs sink to the center of the 
new formed galaxy because of dynamical friction.
In star-dominated environment, N-body simulations 
\citep{bau06,mat07}
have shown that dynamical friction 
is efficient in driving the two MBHs down to a separation at which the enclosed 
stellar mass in the binary is of the order of $m_2$, without
significantly affecting the stellar density profile. This is an indication 
that the evolution is still driven by the dynamical friction exerted by the 
overall distribution of stars, rather than by close individual encounters 
with stars intersecting the binary orbit. 
In our model, we assume that $m_2$ is driven by dynamical friction down 
to a separation $a_0$, where the enclosed stellar mass in the binary is 
twice the mass of the secondary MBH. At that point, we apply the hybrid model
developed by Sesana \citep{sesana10} to follow the semimajor axis and eccentricity
evolution of the MBHB, assuming that the dynamics is purely driven by
interactions with stars, and the stellar density profile
is characterized by a double power law $\rho_r=\rho(r_i) (r/r_i)^{-\gamma}$, with 
$\gamma=2$ for $r>r_i$ and $\gamma=1.5$ for $r<r_i$. Here $r_i=(3-\gamma)G(m_1+m_2)/\sigma^2$ is the 
influence radius of the MBHB, and $\rho(r_i)$ is the stellar density normalization
at $r_i$. The reader is referred to \citep{sesana10} for full details, in the
following we summarize the main features of the model.

On its way to final coalescence starting from $a_0$, the binary is subject 
to three main dynamical mechanisms driving its evolution.

(i) {\it Erosion of the cusp bound to the primary MBH}. In this early 
stage, the MBHB extracts energy and angular momentum from the stars bound to
the primary hole. During this process, lasting $10^5-10^7$yr depending
on the details of the system, the binary shrinks by a factor of ten,
and $e$ usually increases by a large factor, depending on the binary mass 
ratio and on the cusp slope. Results are tabulated in Table 1 of \citep{shm08}.

(ii) {\it Scattering of unbound stars supplied into the binary loss cone by 
relaxation processes}. After the cusp has been depopulated, further hardening
is provided by super-elastic scattering of unbound stars diffused 
into the so called binary loss cone \citep{mm03}. 
In general, hardening by scattering of unbound stars becomes effective 
when the binary reaches the so called hardening radius, defined as \citep{qui96}
$a_h\approx Gm_2/(4\sigma^2)$. This is the separation at which the 
specific binding energy of the binary is of the order of the specific 
kinetic energy of the field stars. Once the binary is hard, 
its hardening proceeds at about 
constant rate and, while circular binary tends to stay circular,  
even slightly eccentric binaries tend to increase their eccentricity 
\citep{qui96,shm06}.
Whether the binary reaches the point at which GW emission becomes
efficient (typically $a_{\rm gw}\lsim 10^{-2}a_h$, see \citep{sesana10}), depends
on the rate at which stars are supplied to the MBHB loss cone. However, as 
noted by \citep{sesana10} the typical eccentricity evolution of the system
depends only mildly on such rate. Here we assume 
efficiently repopulated (also referred to as 'full') loss cone. In this 
case the binary typically enter the GW-dominated phase in a timescale of
$\lsim10^8$yr. 

(iii) {\it Emission of GWs}. 
The effect of GW emission is modeled in the quadrupole 
approximation up to a selected typical GW frequency ($f_{\rm gw}=2f_k$, being $f_k$
the Keplerian frequency of the binary), which is 
2 and $6\times 10^{-5}$ Hz for the high and low seed sources respectively, 
corresponding to orbital separations of many 10's to 100's of $m$.
Under this assumption, the evolution equations
for the system are given by \cite{pm63}
\begin{eqnarray}
\left<\frac{da}{dt}\right> & = & -\frac{64m_1 m_2 m}{a^3(1-e^2)^{7/2}}\left(1+\frac{73}{24}e^2 + \frac{37}{96}e^4\right),\\ \nonumber \\ 
\left<\frac{de}{dt}\right> & = & -e\frac{304 m_1 m_2 m}{a^4 (1-e^2)^{5/2}}\left( 1 + \frac{121}{304}e^2 \right).
\end{eqnarray}
The shrinking rate is a strong factor of $a$, meaning that GW-driven hardening
is effective only at small separations. The eccentricity evolution rate is 
also a strong function of $a$ and $e$ itself, and it is always negative. 
GW emission, therefore, is very effective in circularizing MBHBs, which, in
turn, is the reason why little attention has been paid so far to eccentric
systems in the context of GW detection.

We numerically solve two coupled differential equations for the evolution
of $a$ and $e$, by combining the three dynamical mechanisms 
mentioned above, as detailed in Section 2.3 of \citep{sesana10}. 
We consider three different values of $e_0=0, 0.3, 0.6$, at the initial
semimajor axis $a_0$. This accounts for the fact that galaxies typically
capture each other on very eccentric orbits, which are reflected in the 
initial trajectory of the two MBHs. Even though dynamical friction
against massive gaseous disks has been proven efficient in circularizing the
orbits of the two MBHs \citep{dotti07}, this is in general not true in 
stellar dominated environments \citep{cmg99}, 
and the two MBHs may pair together on a significantly eccentric orbit at $a_0$.
 
\begin{figure*}
\centering
\begin{tabular}{cc}
\includegraphics[width=0.46\linewidth]{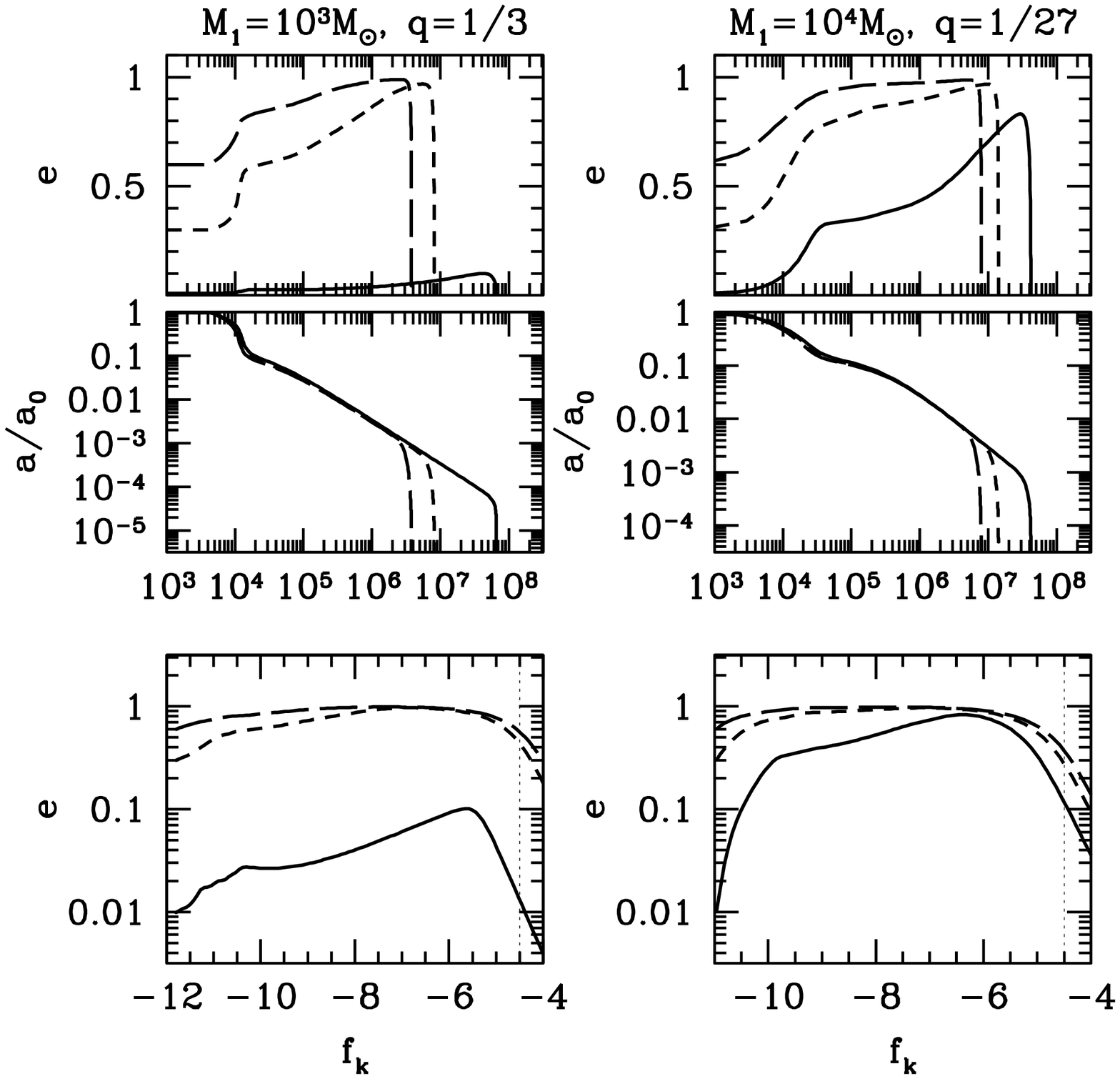} &
\includegraphics[width=0.46\linewidth]{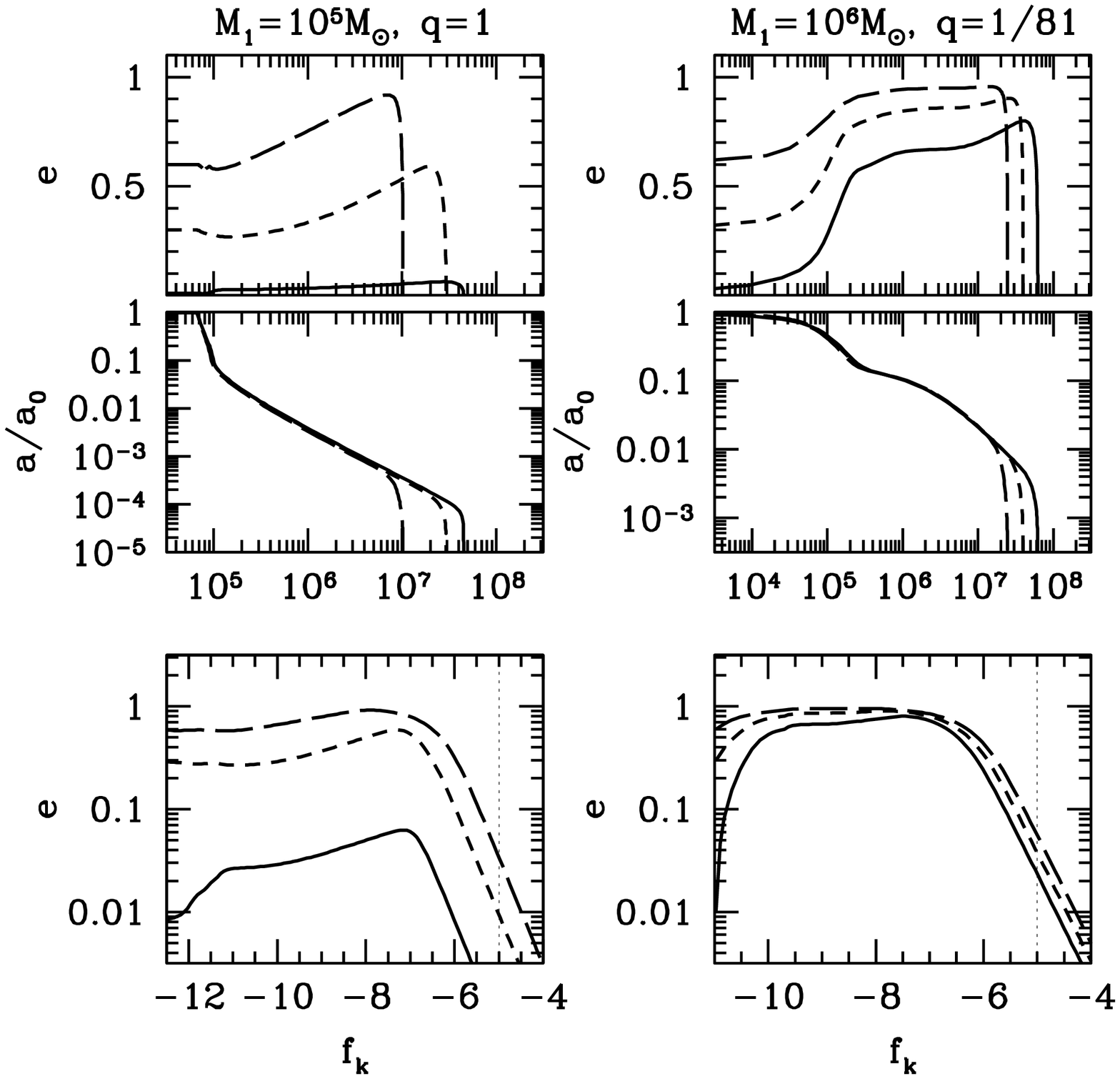} \\
\end{tabular}
\caption{MBHB evolutionary tracks produced by our model for selected
MBHBs. In the top and central panels we plot the evolution of $e$
and $a/a_0$ as a function of time, respectively; in the lower panels
we represent the evolution of $e$ as a function of the Keplerian
frequency $f_k$ of the system. Different line-styles refer to $e_0=0$
(solid), 0.3 (short--dashed), and 0.6 (long--dashed). Dotted vertical
lines in the lower panels mark $f_k=f_{\rm gw}/2=10^{-5}$ and $3\times 10^{-5}$ Hz.}  
\label{tracks}
\end{figure*}

In figure \ref{tracks}, we plot the evolution of the MBHB semi-major axis 
and eccentricity, for values of $m_1$ and $q$ representative of both the light 
and the heavy seed scenarios, for different values of $e_0$. As discussed in 
\citep{sesana10}, when $e_0=0$, 
the eccentricity growth is more significant for unequal mass MBHBs ; while 
binaries with $e_0\gsim 0.3$ reach values of $e$ close to unity irrespective
on the other binary parameters.
Lighter binaries generally preserve higher residual eccentricities in
the {\it LISA} band. The lower panels represent the evolution of
$e$ versus the orbital Keplerian frequency $f_k$. When we start the
PN evolution ($f_k=f_{\rm gw}/2=10^{-5}$ and $3\times 10^{-5}$ Hz for the heavy and 
the light seed models, respectively), MBHB eccentricities are in general
less than $0.1$ for MBHBs with $e_0=0$, but they can be as high
as $\sim0.8$ when $e_0\gsim 0.3$.

We now have six catalogs of MBHBs, three for each seed model,
assuming $e_0=0, 0.3, 0.6$. Each catalog contains a Monte Carlo realization
of the coalescing MBHB population and includes the source redshift $z$, $m_1$,
$m_2$, and $e$ at the selected reference Keplerian frequency.  
 
\begin{figure}[t]
\begin{center}
\epsfig{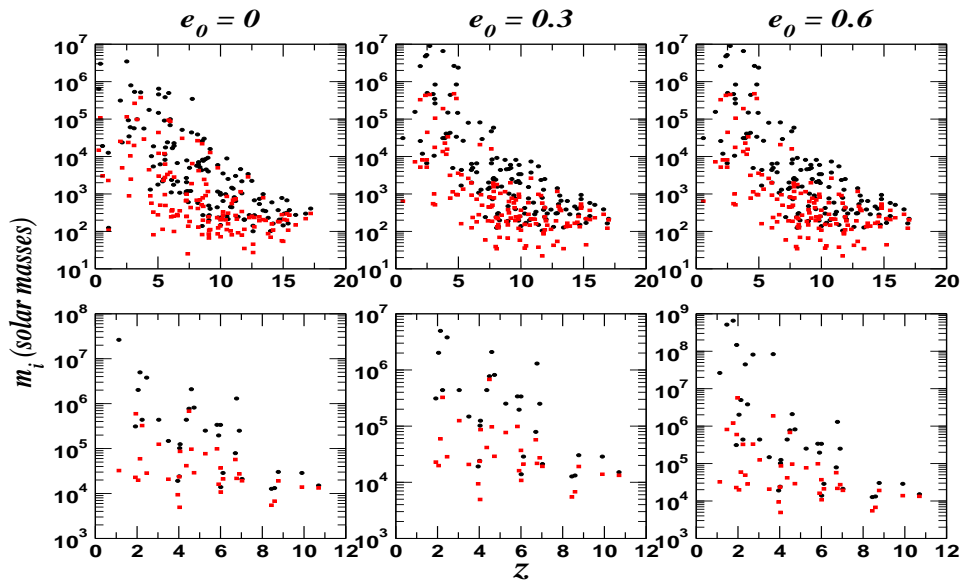}
\end{center}
\caption{The source frame individual mass distributions as a function of redshift for low mass (top) and high mass (bottom) seed black hole binaries.  The three cells going from left to right, both top and bottom, show the catalog individual masses assuming initial eccentricities at binary hardening of $e_0 = 0, 0.3$ and 0.6 respectively.  In all cells $m_1$ are denoted by the black circles, and $m_2$ are represented by the red squares.}
\label{fig:massvsz}
\end{figure}


\begin{figure}
\centering
\epsfig{file=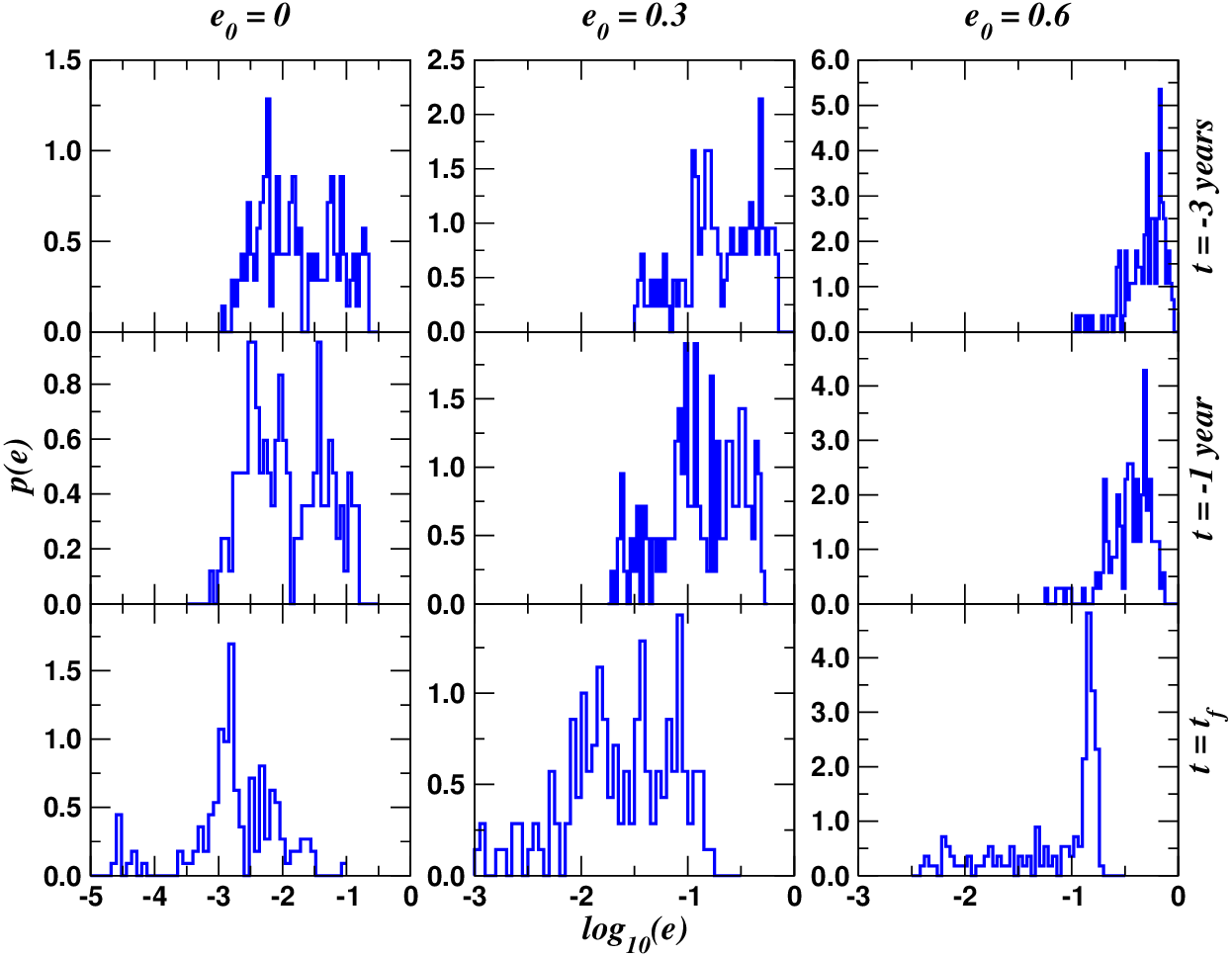, width=5in, height=3in}
\caption{Eccentricity evolution over the final three year period for low mass seed black hole binaries.  The plot displays the evolution for the three models of initial eccentricity.  Note that the evolution reads downwards for each model.}\label{fig:lmseev}
\vspace{1cm}
\epsfig{file=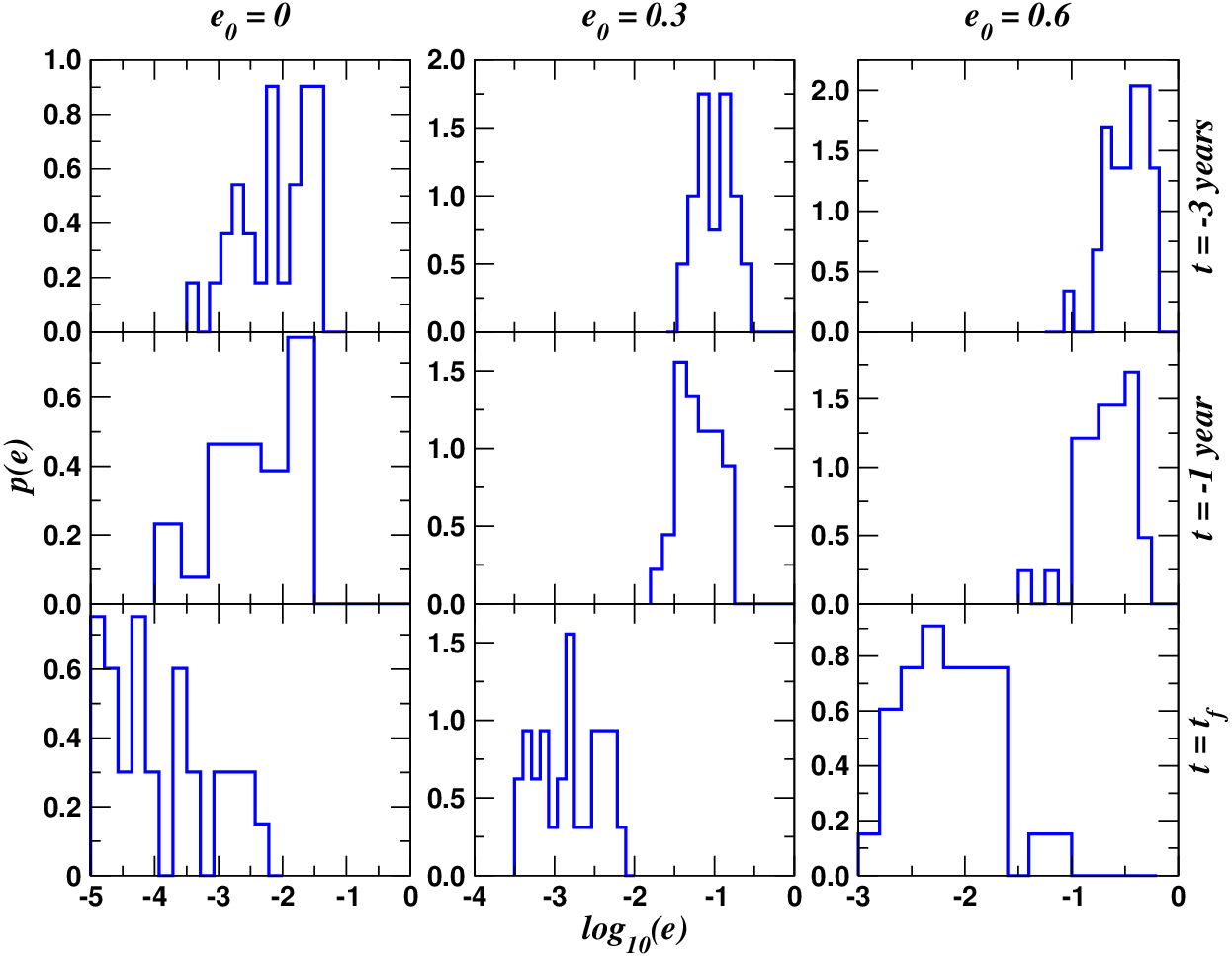, width=5in, height=3in}
\caption{Eccentricity evolution over the final three year period for high mass seed black hole binaries.  The plot displays the evolution for the three models of initial eccentricity.  Note that the evolution reads downwards for each model.}\label{fig:hmseev}
\end{figure}

\section{Comparing Circular and Eccentric Templates for LISA.}\label{sec:mc}
As previously stated, our aim in this study is to evaluate the fidelity of restricted PN circular inspiral templates in the presence of eccentric SMBHBs.  A full exploration of eccentric black hole binary parameter estimation is currently underway~\cite{porses}, but is outside the scope of this particular study.  To compare the template families, we ran a  Monte Carlo simulation where we assumed a one year LISA mission lifetime.  The main reason for this is, most of the SNR comes from the end of the inspiral.  Therefore, as the starting point of observation is arbitrary, it is our ability to match the eccentric waveform for a coalescing binary in the final year that is important.  We  also investigate the low and high mass seed cases individually.  We will explain the organization of the Monte Carlo runs in greater detail below.

\subsection{The Monte Carlo Setup.}
The first point of importance is termination of the eccentric inspiral waveforms.  As we do not know the exact position of the last stable orbit (LSO) for eccentric binaries, we decided to terminate the waveforms before the orbital separation reached $r=6m$ (i.e. the LSO for a test particle in a Schwarzschild geometry) or 5 mHz, whichever is reached first.  In Figures~(\ref{fig:lmseev}) and~(\ref{fig:hmseev}), we plot the evolution of eccentricity at -3 years, -1 year and at the final point of evolution.   These catalogues provide the initial conditions in terms of eccentricity for our simulations.

The Monte Carlo was set up as follows : We started with six source catalogues consisting of high or low mass seeds with initial eccentricities of $e_0 = 0, 0.3$ or 0.6 at binary hardening.  For each of the sources in the eccentric catalogues, we randomized the sources over the parameters $(\iota, \psi, \varphi_c, \theta, \phi)$, while using the catalogue values of $(m_1, m_2, z, x_0, e_0)$.  For each system,  we calculated the optimal signal to noise ratio (SNR) to ensure we achieve a threshold value of 8.  This value was chosen as it is a value at which many algorithms start to discern between a real SMBHB signal and a signal produced by the multitude of white dwarf binaries in the galactic foreground.  The optimal SNR is defined as
\begin{equation}
\rho_{opt} = \left<h_e \left | \right . h_e\right> ^{1/2},
\end{equation}
where $h_e$ denotes an eccentric template, and the angular brackets represent the noise weighted inner product
\begin{equation}
\left<a \left |  b\right> \right .= 2\int_{f_0}^{f_h}\,\frac{df}{S_n(f)}\, \tilde{a}(f)\tilde{b}^{*}(f) + cc,
\end{equation}
where 
\begin{equation}
\tilde{a}(f) =  \int_{-\infty}^{\infty}\, dt\, a(t)e^{2\pi\imath ft},
\end{equation}
is the Fourier transform of the time domain function $a(t)$ and the $cc$ denotes complex conjugate.  The integration limits $(f_0, f_h)$ are given by $f_0 = 10^{-5}$ Hz,
 and $f_h = f_{gw}(r = 6m)$ or $5\times 10^{-3}$ Hz, depending on the source.  The quantity $S_n(f)$ is the one-sided power spectral density and is composed of a combination of both instrumental noise and the confusion limited galactic background, i.e.
 \begin{equation}
 S_n(f) = S_n^{\rm instr}(f) + S_n^{\rm conf}(f),
 \end{equation}
 where the instrumental noise $S_n^{\rm instr}(f)$ is given by
\begin{eqnarray}
S_{n}^{\rm instr}(f) &= &\frac{1}{4L^{2}}\left[ 2S_{n}^{pos}(f) \left(2+\left(\frac{f}{f_{*}}\right)\right) + 8 S_{n}^{\rm acc}(f)  \left(1+\cos^2\left(\frac{f}{f_{*}}\right)\right)  \left( \frac{1}{(2\pi f)^{4}} + \frac{(2\pi 10^{-4})^{2}}{(2\pi f)^{6}}   \right)   \right] , \nonumber
\end{eqnarray}
 and the confusion noise estimate $S_{n}^{\rm conf}(f)$ is derived from a Nelemans, Yungelson, Zwart galactic foreground model~(\cite{NYZ, TRC})
\begin{equation}
S_{n}^{\rm conf}(f) = \left\{ \begin{array}{ll} 10^{-44.62}f^{-2.3} & 10^{-4} < f\leq 10^{-3} \\ \\ 10^{-50.92}f^{-4.4} & 10^{-3} < f\leq 10^{-2.7}\\ \\ 10^{-62.8}f^{-8.8} &  10^{-2.7} < f\leq 10^{-2.4}\\ \\ 10^{-89.68}f^{-20} &  10^{-2.4} < f\leq 10^{-2}  \end{array}\right. .
\end{equation}
In the above expressions,  $L=5\times10^{6}$ km is the arm-length for LISA,  $S_{n}^{pos}(f) = 4\times10^{-22}\,m^{2}/Hz$ and $S_{n}^{\rm acc}(f) = 9\times10^{-30}\,m^{2}/s^{4}/Hz$ are the position and acceleration noises respectively.  The quantity $f_{*}=1/(2\pi L)$ is the mean transfer frequency for the LISA arm.   The instrumental noise also contains a reddened noise term which steepens the noise curve between $10^{-4}$ and $10^{-5}$ Hz.  Finally, the units of $S_{n}^{\rm conf}(f)$ are Hz$^{-1}$.

 If the optimal SNR is greater than the threshold  of 8, we then run a short Markov Chain Monte Carlo (MCMC) with circular templates, using the exact parameters of the eccentric binary system as our initial best guess for the circular templates.  The MCMC is a stochastic search method which has been used many times in LISA GW data analysis.  While these works used sophisticated variants of the MCMC to search over a
 wide parameter space, in this study we are only concerned with local exploration and are therefore using a straightforward MCMC.  While we refer the reader to~\cite{cp1} for an indepth discussion about MCMC methodology in GW data analysis, briefly, the method works as follows : starting with the signal $s(t) = h_e(t) + n(t)$, where $n(t)$ is the noise in the detector, and some initial circular template $h_c(t;\vec{\lambda}_i)$ constructed by choosing a random starting point in the parameter space $\vec{\lambda}_i$, we then draw from a proposal distribution and propose a jump to another point in the space $\vec{\lambda}_{i+1}$.  In order to compare both points, we evaluate the Metropolis-Hastings ratio
\begin{equation}
H = \frac{\pi(\vec{\lambda}_{i+1})p(s|\vec{\lambda}_{i+1})q(\vec{\lambda}_i|\vec{\lambda}_{i+1})}{\pi(\vec{\lambda}_i)p(s|\vec{\lambda}_i)q(\vec{\lambda}_{i+1}|\vec{\lambda}_i)}.
\end{equation}
Here $\pi(\vec{\lambda}_i)$ are the priors of the parameters and $p(s|\vec{\lambda}_i)$ is the likelihood defined by
\begin{equation}\label{eqn:likelihood}
{\mathcal L}\left(\vec{\lambda}_i\right) = C\,e^{-\left<s-h_c\left(\vec{\lambda}_i\right)|s-h_c\left(\vec{\lambda}_i\right)\right>/2},  
\end{equation}
where $C$ is a normalization constant.  The quantity $q(\vec{\lambda}_i|\vec{y})$ is the proposal distribution used for jumping from $\vec{\lambda}_i$ to $\vec{\lambda}_{i+1}$.  For this study, the proposal distribution is a multivariate Gaussian calculated using the Fisher information matrix (FIM)
\begin{equation}
\Gamma_{\mu\nu} = \left<\frac{\partial h_c}{\partial \lambda^{\mu}} \left |  \frac{\partial h_c}{\partial \lambda^{\nu}}\right> \right . .
\end{equation}
Once a jump is proposed, it is then accepted with probability $\alpha = min(1,H)$, otherwise the chain stays at $\vec{\lambda}_i$.   

In order to speed up the convergence of a Markov chain, it has been shown~\cite{cp1} that heating the likelihood surface via simulated annealing helps the chain to move more easily by effectively smoothening and reducing the height of maxima on the surface.  To this end, we run the first 1000 iterations of the MCMC using a simulated annealing phase which replaces the value of 1/2 in the exponent of Eqn~(\ref{eqn:likelihood}) by a factor $\beta$ where
\begin{equation}
\beta = \left\{ \begin{array}{ll} \frac{1}{2}10^{-\xi\left(1-\frac{i}{T_{c}}\right)} & 0\leq i\leq T_{c} \\ \\ \frac{1}{2} & i > T_{c}  \end{array}\right. .
\end{equation}
For this particular study,  $\xi$ is the heat-index defining the initial heat and is taken to be equivalent to the optimal SNR calculated using the circular templates, $i $ is the number of steps in the chain and $T_{c} = 1000$ is the cooling schedule.  We should remark here on our choice of the initial heat.  We tried scaling the heat to the optimal SNR for eccentric binaries, but this make the initial heat very high and as we start the MCMC close to the true solution, this amounts to a wasted number of computer cycles.  As in general, the optimal SNR of a circular binary is less than the optimal SNR of an eccentric binary (something we will justify at a later point), we use the circular binary SNR to scale the initial heat.

One of the quantities that we investigate is the overlap between the template and the signal.  While the concept of a global overlap for LISA is ill-defined, the overlap in a particular channel is defined by
\begin{equation}
{\mathcal O} = \frac{ \left<h_c \left | \right . h_e\right> }{\sqrt{ \left<h_c \left | \right . h_c\right>  \left<h_e \left | \right . h_e\right> }}.
\end{equation}
While the MCMC changes the parameters of the system, the overlap is a simple way of measuring the improvement in the fit between the eccentric and circular templates.
 
 \begin{figure}
\centering
\epsfig{file=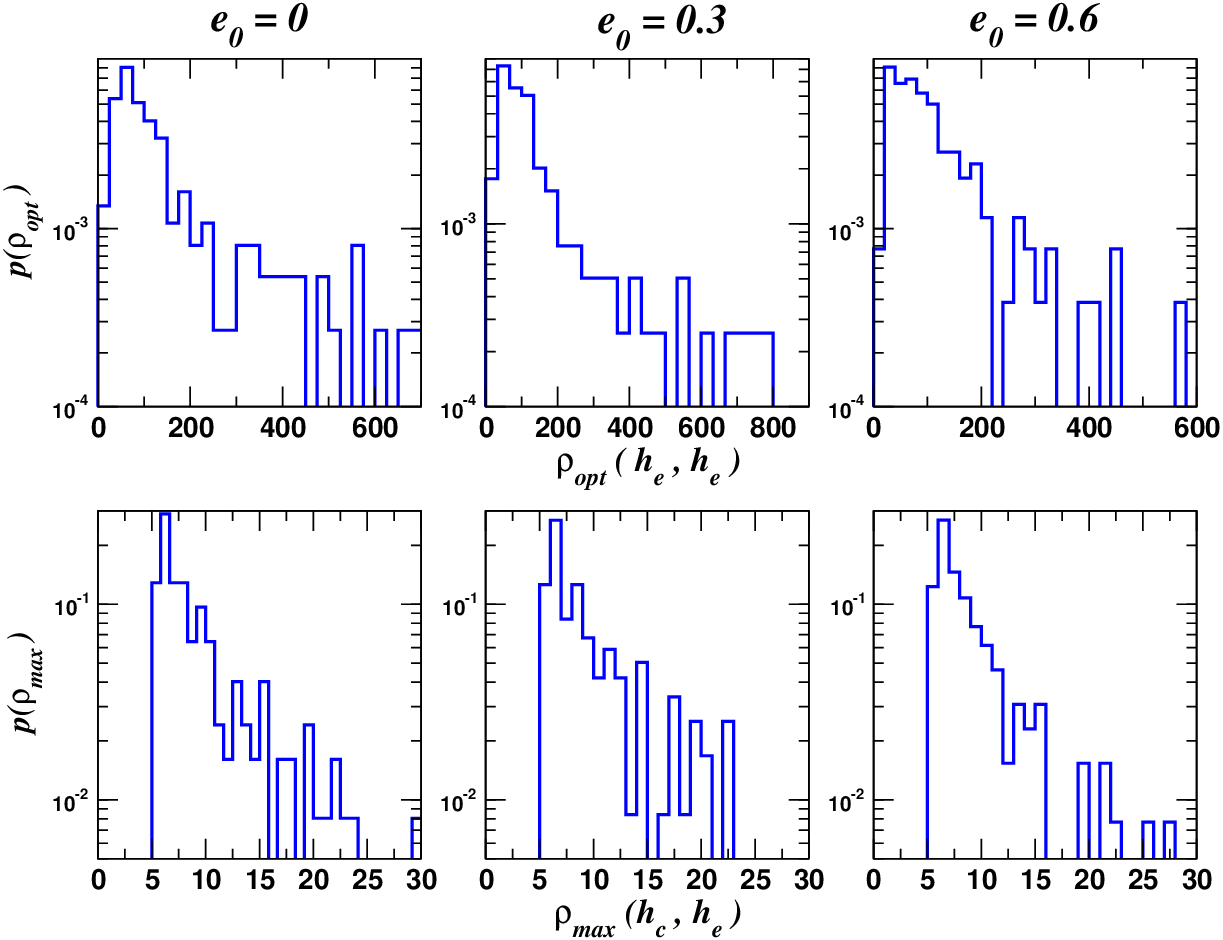, width=5in, height=3in}
\caption{The top row of this figure shows the optimal SNR distribution using high mass seed eccentric waveforms for each of the three initial eccentricities.  The bottom row shows the maximum SNRs recovered by the MCMC using circular templates.  We can see a massive degradation in the recovered SNR using circular templates.}\label{fig:hssnr}
\vspace{1cm}
\epsfig{file=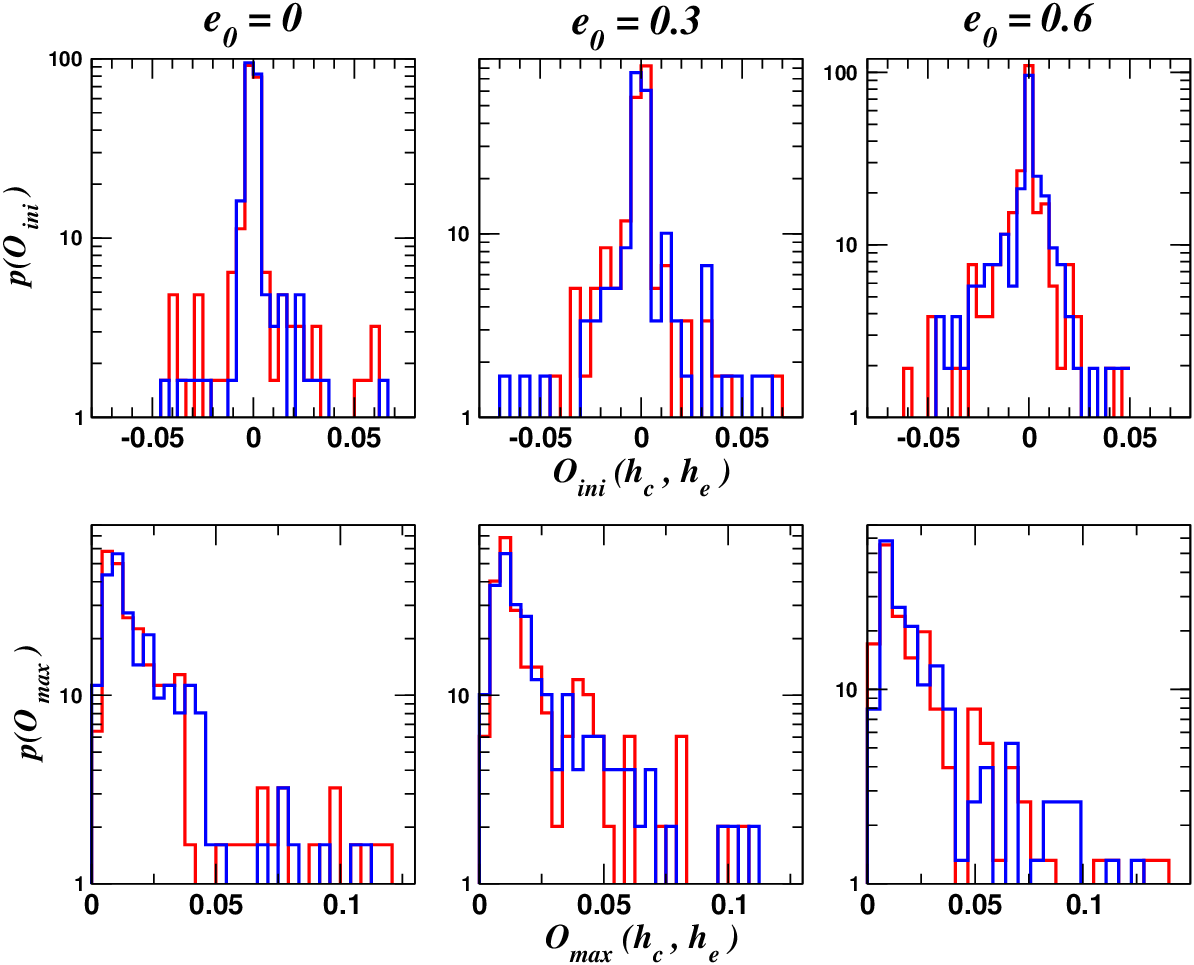, width=5in, height=3in}
\caption{The top row of this figure shows the initial distribution of overlaps achieved by comparing circular and eccentric templates for high mass seeds with exactly the same parameters for each of the three initial eccentricities in the LISA A (blue) and E (red) channels.  The bottom row shows the distribution of overlaps at the end of the MCMC.  While there is an improvement in fit, the overlaps are not close to what we require for LISA.}\label{fig:hsol}
\end{figure}

\section{Results.}\label{sec:results}
In the following sections we present the results for the 1 year missions for both the low and high mass seeds seperately.
\subsection{High Mass Seed Black Hole Binaries}
In Fig~(\ref{fig:hssnr}) we plot both the optimal and maximum recovered SNRs from the high mass seed Monte Carlo's.  We can see from the top panels that optimal SNRs for eccentric binaries peak at values of 50-100, with varying maximal SNRs of between 600 and 800.  In the bottom panels we have plotted the maximum SNR recovered by a maximized circular template at the end of the MCMC.  We can see that for all eccentric catalogues, the distributions peak at SNRs of 5-10, and have maximum values of 25-30.  This means that for the high mass seed systems, the circular templates have median optimal SNR recovery of just 5, 3 and 5\% for the three models of initial eccentricity.

To properly explain this result, we refer to Fig~(\ref{fig:hsol}).  In the top panels we plot the initial overlap between a circular and eccentric waveform with exactly the same parameters in both the A (blue) and E (red) LISA channels.  We can see that while the distributions vary from slightly negative to slightly positive, they are roughly peaked around zero.  This implies that circular and eccentric templates with exactly the same parameters are essentially orthogonal to each other, regardless of the value of the initial eccentricity.  In the bottom panels we show the distributions of maximum overlaps at the end of the MCMC.  It is clear that the MCMC has managed in each case to find a more suitable parameter set in terms of improving the likelihood between the circular and eccentric templates, but the overlaps are no-where near what would be required for LISA data analysis.  The best that we could do with the circular templates is confirm a detection, as long as the threshold for a detection is sufficiently high.

\begin{figure}
\epsfig{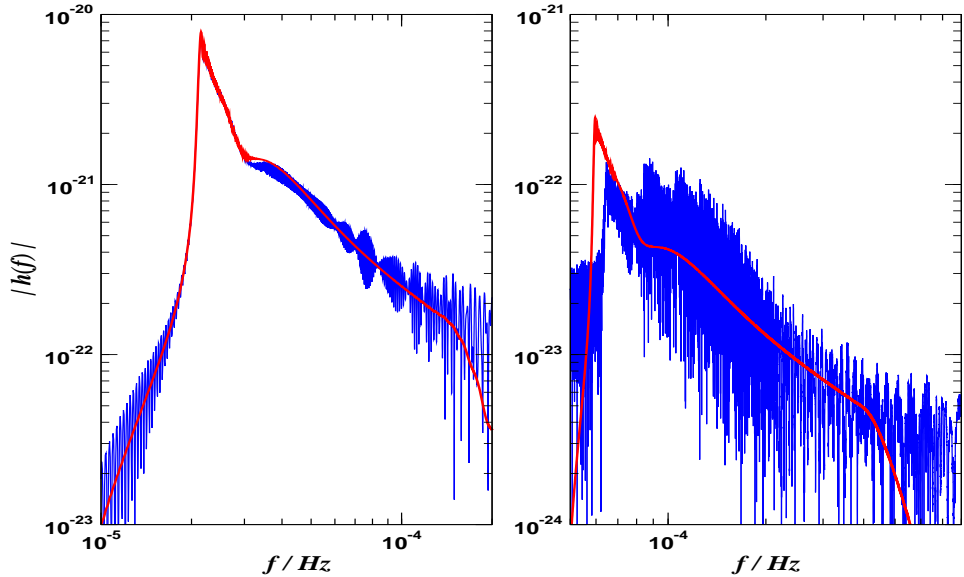}
\caption{A comparison of the spectra of two circular and eccentric systems with identical paramters.  The eccentric system on the left corresponds to a high mass seed system with initial and final eccentricities of $1.8\times10^{-4}$ and $2.3\times10^{-5}$, while the system on the right has initial and final eccentricities of 0.438 and 0.011 .   We can see that in both cases the circular templates are not a good match to the eccentric waveforms.}\label{fig:wspec}
\end{figure}

To further explain the disparity between circular and eccentric templates, we have plotted the spectra of a some SMBHB systems in Fig~(\ref{fig:wspec}).  The system in the left hand panel has individual redshifted masses of $4.238\times10^6$ and $3.728\times10^6\,M_{\odot}$ at $z=4.49$.  For this system the initial and final eccentricities are $e_i=1.8\times10^{-4}$ and $e_f=2.3\times10^{-5}$.   The system in the right hand panel has individual redshifted masses of $1.975\times10^6$ and $2.157\times10^5\,M_{\odot}$ at $z=6.9$.  For this system the initial (this we define as being the eccentricity at the beginning of the observation) and final (when the binary separation reaches $6m$ or 5 mHz) eccentricities are $e_i=0.438$ and $e_f=0.011$.  Both systems have the same angular parameters and sky locations.  If we first focus on the system in the left hand cell, while it looks like there is a good match between the eccentric (blue) and circular (red) templates at low frequencies (we should point out here that if we zoom in on these frequencies, there is a clear phase mismatch between the waveforms), it is clear that even for an extremely mildly eccentric binary, there is extra power at higher frequencies that we do not see in circular templates.  As quantities such as overlaps and SNRs require good phase matching, we can now see why circular templates do such a bad job in capturing eccentric binaries for LISA.  In the panel on the right, for an eccentric waveform with substantial eccentricity, it is clear that there is very little possibility of a circular template being able to match the higher power and frequency content of such an eccentric binary.  Even if an algorithm managed to find a system with low enough masses that it pushed the LSO frequency close to that of the eccentric binary, it would still not be able to capture the structural information. 

\begin{figure}
\epsfig{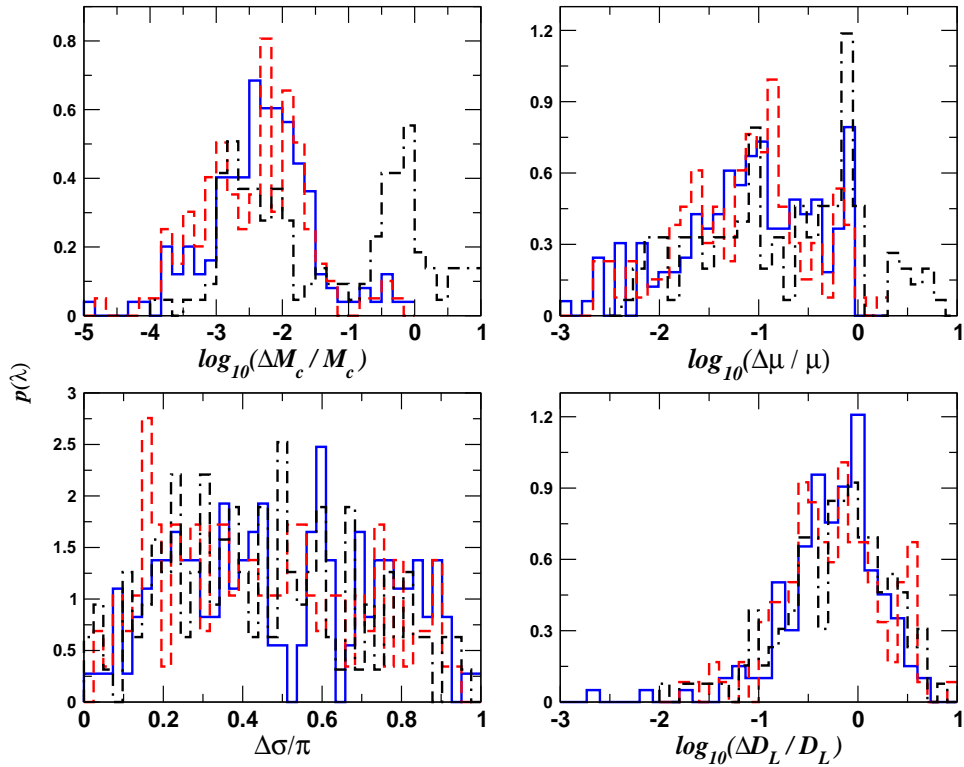}
\caption{High mass seed black hole parameter mismatch for chirp-mass, reduced mass, sky position and luminosity distance for the three models with initial eccentricity of $e_0=0$ (solid line), $e_0 = 0.3$ (dashed line) and $e_0 = 0.6$ (dot-dashed line).}\label{fig:hmspm}
\end{figure}

While it will not be possible to carry out a parameter recovery of eccentric binaries using circular templates, as we said earlier, it may still be possible to use the circular templates in a detection only framework.  Thus, it is still interesting to investigate the parameter mismatch between the two waveform families.  In Fig.~(\ref{fig:hmspm}) we plot the parameter mismatch at the end of the MCMC for chirp-mass, reduced mass, luminosity distance and sky position.  For the sky position error $\Delta\sigma$, we calculate the orthodromic distance between the sky position of the eccentric binary and the sky position of the best fit circular template.  This is done using a special case of the Vincenty formula (which is normally used for calculating the distance between two points on an ellipsoid):
\begin{equation}
\Delta\sigma = arctan\left(\frac{\sqrt{\left(\cos\phi_r\sin\Delta\theta_L\right)^2 + \left(\cos\phi_a\sin\phi_r-\sin\phi_a\cos\phi_r\cos\Delta\theta_L\right)^2}}{\sin\phi_a\sin\phi_r+\cos\phi_a\cos\phi_r\cos\Delta\theta_L}\right),
\end{equation}
where $\Delta\sigma$ has units of radians, $(\phi_a, \phi_r)$ denote the actual and recovered longitudes and $\Delta\theta_L = \theta_L^a - \theta_L^r$ is the difference in actual and recovered latitudes.  We have used this particular expression as to avoid the large rounding errors associated with the spherical law of cosines for the case where $\Delta\sigma\ll 1$ and with the haversine formula in the case of antipodal points.

We can see from Fig~(\ref{fig:hmspm}) that for all three initial eccentricity models, the errors in the parameter estimation are bigger than we are used to for these types of sources~\cite{cp1}.  If we first focus on the mass parameters, for the models where the initial eccentricity was 0 or 0.3, we can see that the chirp and reduced mass fractional error distributions are peaked at approximately $10^{-2}$ and $10^{-1}$ respectively.  As a comparison, MCMC searches have recovered fractional errors in both mass parameters on the orders of $10^{-6}$ and $10^{-4}$ for circular SMBHBs~\cite{cp1}.  For the $e_0 = 0.6$ case, we can see that while a number of the sources are resolvable with similar precision as in the other two cases, there are a number of sources where the mass parameters are essentially undetermined.  Now focusing on the error in the sky position, for all three cases, we can see that the sky is essentially undetermined with massive errors in the final estimated sky position.  Finally, for the estimation of luminosity distance, we end up in a situation where most of the distances are unresolved.  The overwhelming conclusion here is, while for the higher SNR cases we may have confidence in a detection, we can not have confidence in the system parameters extracted using circular templates.

\subsection{Low Mass Seed Black Hole Binaries.}
\begin{figure}
\centering
\epsfig{file=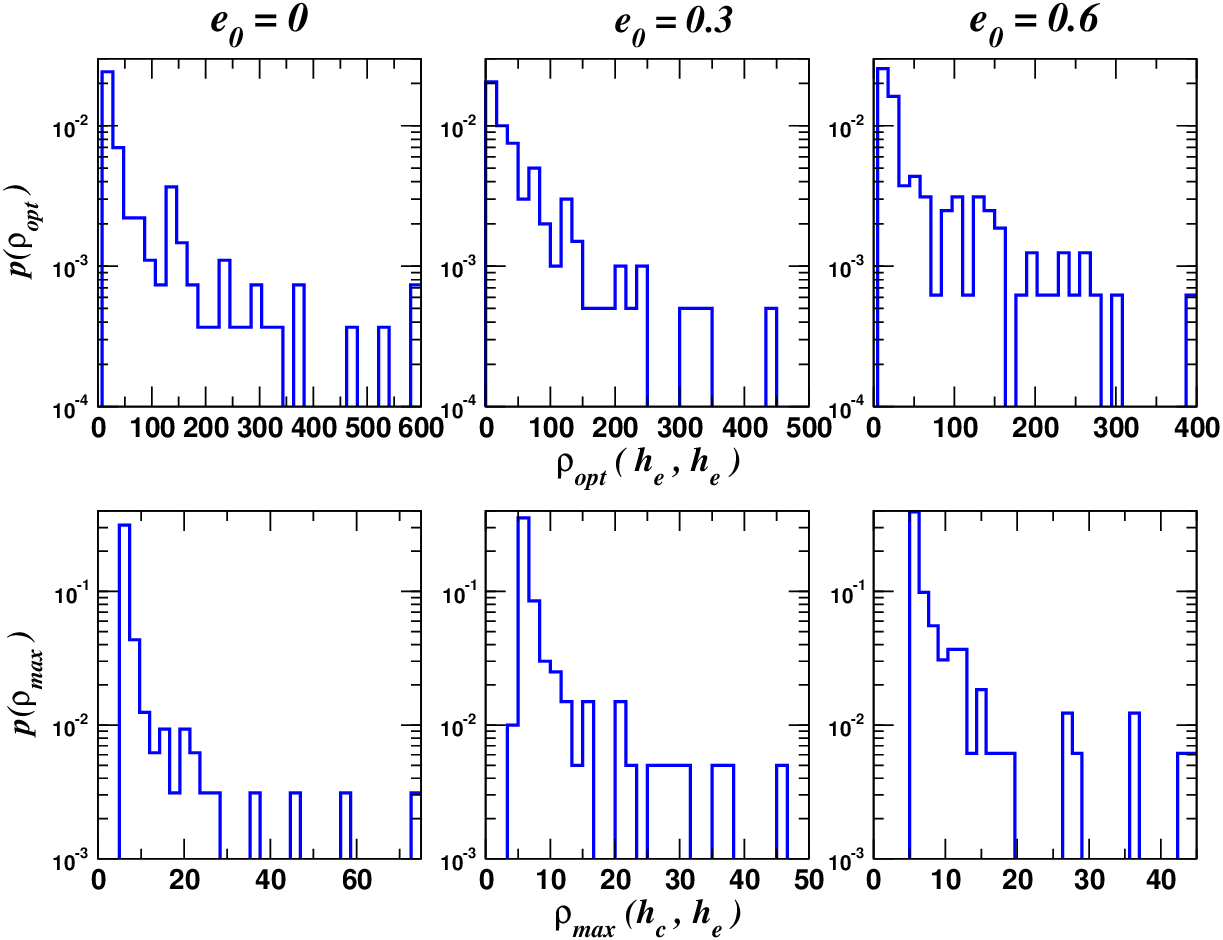, width=5in, height=3in}
\caption{The top row of this figure shows the optimal SNR distribution using low mass seed eccentric waveforms for each of the three initial eccentricities.  The bottom row shows the maximum SNRs recovered by the MCMC using circular templates.  We can see a massive degradation in the recovered SNR using circular templates.}\label{fig:lmsnr}
\vspace{1cm}
\epsfig{file=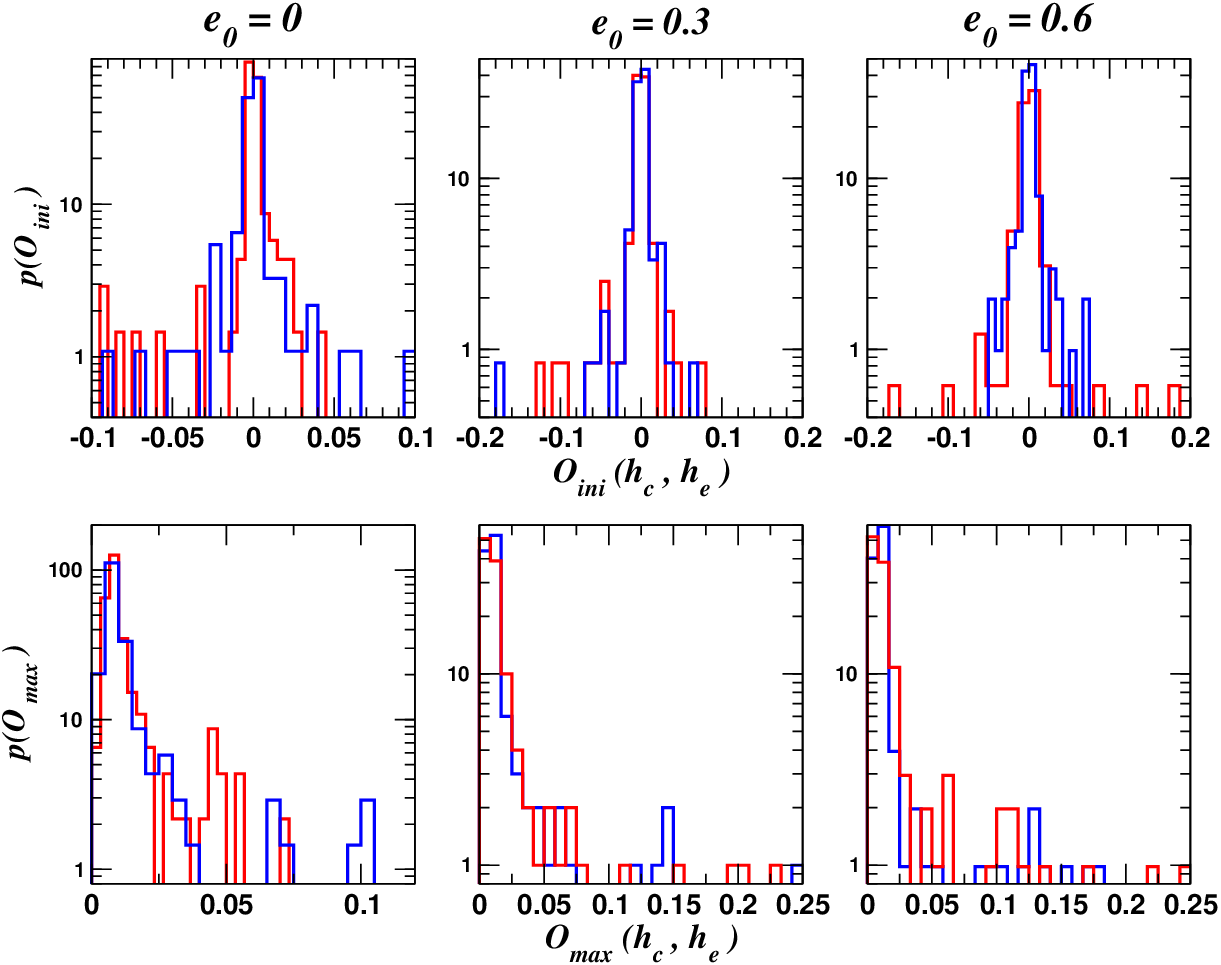, width=5in, height=3in}
\caption{The top row of this figure shows the initial distribution of overlaps achieved by comparing circular and eccentric templates for low mass seeds with exactly the same parameters for each of the three initial eccentricities in the LISA A (blue) and E (red) channels.  The bottom row shows the distribution of overlaps at the end of the MCMC.  While there is an improvement in fit, the overlaps are not close to what we require for LISA.}\label{fig:lsol}
\end{figure}

For low mass seed SMBHBs, the situation is slightly better.  There are a few reasons for this : as we previously stated, if a system does not reach the orbital separation of $r=6m$ before reaching a GW frequency of 5 mHz, we terminate the waveform at 5 mHz.  Because of the masses involved, the coalescence frequencies of many of the systems are at 10's of mHz and are thus outside of our band of interest for this study.  This effect can be seen in the optimal SNRs presented in the top panels of Fig~(\ref{fig:lmsnr}).  While we again have SNRs of 400-600, the distributions are peaked at SNRs of 10-20.  We know from previous studies that most of the SNR is recovered from the final few cycles, usually corresponding to the last few days of inspiral.  As we do not see the coalescences for most of these systems, the circular templates have to fit a less relativistic waveform and thus have the ability to perform better.  Also, we can see from Figs~(\ref{fig:lmseev}) and (\ref{fig:hmseev}) that the eccentricity evolution is slower for the low mass seeds.  This means that while the eccentricity of low mass seeds stays higher for a greater period of time, there is more possibility for the circular template to obtain a better fit as the system is not changing as quickly as the high mass seed case.  Finally, we remarked earlier that eccentric binaries radiate at higher harmonics of the orbital phase.  In the low mass seed case, as we are truncated virtually all of the systems at 5 mHz, the effect of the radiation at higher harmonics is not as influential as in the high mass seed case (we do note, of course, that this is due to the use mainly to our use of the LFA for the LISA response).  This effect has also been seen in the case of circular binaries where higher harmonic corrections have been added~\cite{pc}.We can see from the bottom panels of Fig~(\ref{fig:lmsnr}) that while the circular templates still suffer, they actually recover a greater percentage of the optimal SNR than in the high mass seed case.  While the SNR distributions are still peaking at low values of 5-10, we are seeing systems with maximum recovered SNRs of between 40 and 80.  For the low mass seeds, we are now achieving median optimal SNR recoveries of 13, 10 and 12\% respectively for the three models of initial eccentricity.

If we now move our attention to Fig~(\ref{fig:lsol}), we again plot the initial overlaps for circular and eccentric templates with identical parameters for the LISA A (blue) and E (red) channels in the top panels.  We see a similar story here to the high mass seed case where the initial overlaps are peaked around zero, again showing that the two template families with identical parameters are essentially orthogonal to each other.  In the bottom panels we plot the maximum overlaps at the end of the MCMC.  Again, while the peak of the distributions is still close to zero, we do see systems with overlaps approaching 0.25, showing an improvement over the high mass seed case.  We should once more point out that with identical parameters, the initial overlaps seem to be independent of the initial eccentricity.

\begin{figure}
\epsfig{file=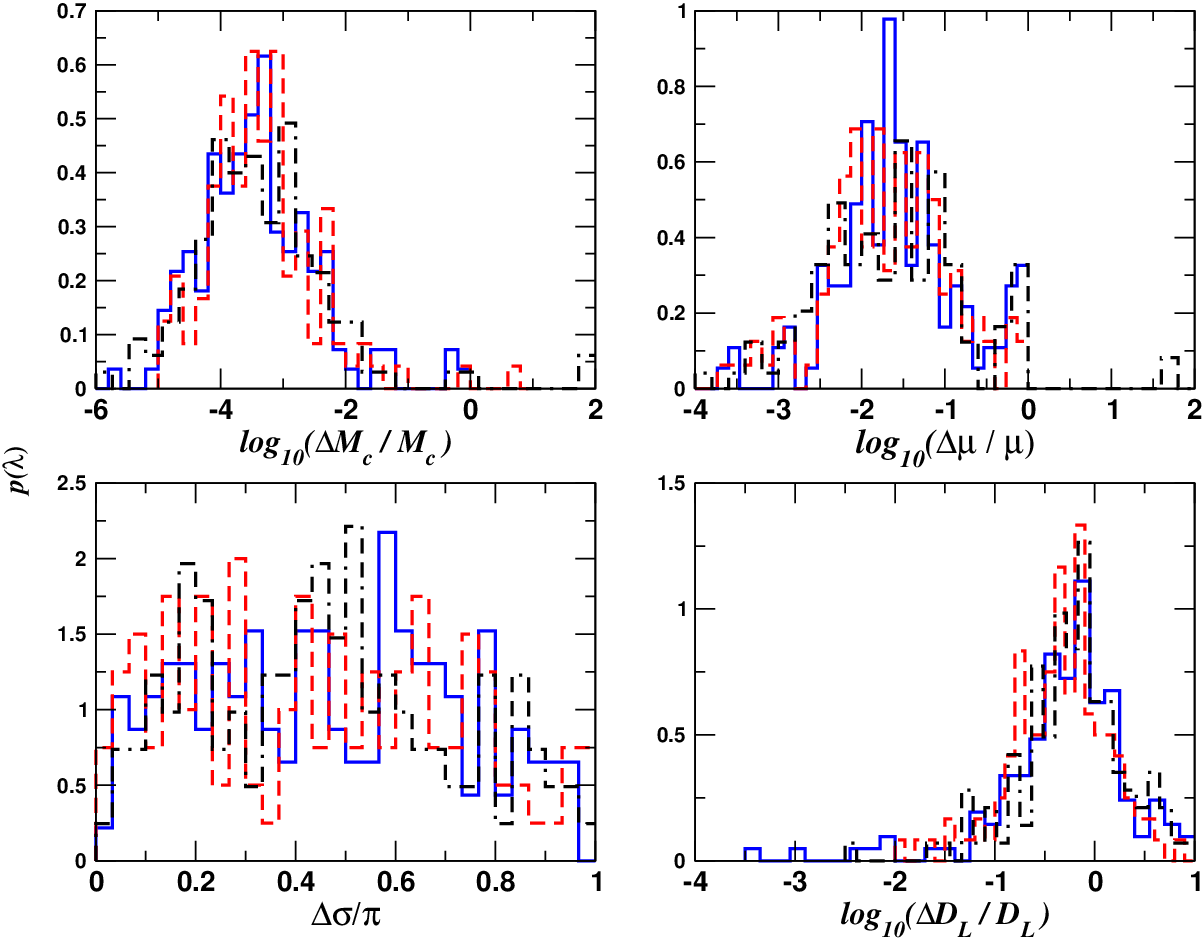, width=5in, height=4in}
\caption{Low mass seed parameter mismatch for chirp-mass, reduced mass, sky position and luminosity distance for the three models with initial eccentricity of $e_0=0$ (solid line), $e_0 = 0.3$ (dashed line) and $e_0 = 0.6$ (dot-dashed line).}\label{fig:lmspm}
\end{figure}

While a maximum overlap of 0.25 is an improvement, it would not really give us enough confidence in our detection.  However, just as in the high mass seed case, it is interesting to look at the effect of parameter estimation in the low mass seed case.  In Fig~(\ref{fig:lmspm}) we plot the errors in parameter estimation for the chirp and reduced masses, the sky position and the luminosity distance.  It is here that we see the benefit of not having to fit the merger of the waveform.  We can now see that for all three models, the fractional errors in the chirp mass estimate peak somewhere between $10^{-4}$ and $10^{-3}$.  For the reduced mass, while some systems are unresolvable in this parameter, the vast majority peak with errors between $10^{-2}$ and $10^{-1}$.  However, this is where the good news ends.  We again see that in all three models both the sky position and luminosity distance are essentially unresolved for all systems.  So once again, the conclusion is that also in the case of low mass seeds, circular templates are not efficient enough in capturing eccentric black hole binaries.

\section{Conclusion.}
In this work, we have used a number of source catalogues taken from the end stage of a hybrid model for the evolution of eccentric SMBHBs.  These catalogues
describe both high and low mass seed systems with initial model eccentricities of 0, 0.3 and 0.6.  These systems were then evolved into the LISA detection window
using the Peters and Mathews' equations to model the secular decay of the semi-major axis $a$ and the eccentricity $e$ due to GW emission.  At a certain point, we then
changed over to the 2-PN equations for the evolution of the PN velocity parameter $x$ and eccentricity.  This evolution then provided us with the initial conditions to examine
the sources in the final year of evolution, either to coalescence or a maximum GW frequency of 5 mHz.

An extensive study was then carried out by combining a Monte Carlo simulation over the extrinsic parameters of the system, combined with a Markov chain Monte Carlo, to examine the fidelity of searching for eccentric systems using circular templates.  We found that in the high mass seed case, only about 5\% of the optimal SNR was recovered by the circular templates.  Worse still was that the fit between the eccentric and circular waveforms at the end of the MCMC only achieved overlaps of about 0.15, much below the confidence level needed for LISA data analysis.  We also looked at the errors in parameter estimation and found that while the mass parameters were resolvable, the errors were quite large.   For all three initial eccentricity models, both the sky position and luminosity distance were unresolvable.  This has the consequence that as the redshifted masses are similar to the true redshifted masses, using a particular cosmological model, we would interpret the true system as being composed of much lighter binaries at a higher redshift.

For the low mass systems, the recovered optimal SNR was better with a maximum of 13\%.  But again, the maximum overlaps were only on the order of 0.25.  While this represented an improvement over the high mass seed case, and can be attributed to the fact that for the vast majority of these systems we did not see the coalescence, thus meaning that the circular templates had to fit a less relativistic waveform, it is still far below the LISA confidence level.  While the errors in the estimation of both the chirp and reduced masses were smaller in this case, the sky position and luminosity distance were again unresolved.

The clear consequence of this study is that LISA data analysis will require the use of eccentric templates, even if the final eccentricity is on the order of $10^{-5}\leq e_f\leq 10^{-4}$.  While ground based studies have demonstrated that circular templates work at this level, the case for LISA is very different.  It was found for LIGO that fitting factors were reduced when neutron star binaries were considered, rather than black hole binaries.  This was due to the fact that a stellar mass black hole binary lasts less than one second in the detector, whereas a neutron star binary is observable for a little over 20 seconds.  In the LISA case, with signals lasting many months or years, there is much more scope for the circular and eccentric templates to be out of phase with each other due to the total number of cycles, so our overall result is really not surprising.  

Our study also has consequences for Numerical Relativity as a lot of work has gone into reducing the residual eccentricity in merger waveforms.  If it is true that eccentric binaries will be observable in LISA, with some of the high eccentricities seen in this study, then we will require a catalogue of merger waveforms for eccentric binaries before LISA launches.

We have now begun two further studies regarding eccentric binaries : the first is an extensive parameter estimation study.  In the second study, we intend to revisit the 
current problem but using circular binaries with higher harmonic corrections.

\begin{acknowledgments}
The authors would like to thank Luciano Rezzolla for interesting discussions.
\end{acknowledgments}


\end{document}